\documentclass[11pt,english]{article}
\usepackage[margin=2cm]{geometry}
\usepackage{comment}
\usepackage[english]{babel}
\usepackage{graphicx}
\usepackage{natbib}
\usepackage{amssymb}
\usepackage{subfigure}
\usepackage{float}
\usepackage{tabularx}
\usepackage{multirow}
\usepackage{lineno}

\usepackage{color, colortbl}
\usepackage{amsmath}
\usepackage{amsfonts}
\usepackage{array}
\usepackage{url}
\usepackage{booktabs}
\definecolor{Gray}{gray}{0.88}
\definecolor{DarkGray}{gray}{0.7}

\newcolumntype{x}[1]{%
>{\centering\arraybackslash\hspace{0pt}}p{#1}}%

\newcommand{\bs}[1]{\boldsymbol{#1}}
\newcommand{\mc}[1]{\mathcal{#1}}
\newcommand{\real}{\mathbb{R}}

\newcommand{\pderiv}[2]{\frac{\partial #1}{\partial #2}}
\newcommand{\pderivsq}[1]{\frac{\partial^2}{(\partial #1)^2}}
\newcommand{\bl}{\begin{linenomath}} 
\newcommand{\el}{\end{linenomath}}

\begin{document}

\bibliographystyle{ams}

\thispagestyle{empty}

\vspace{8mm}

\begin{center}

\vspace{6mm}

{\LARGE Bayesian hierarchical modeling of extreme hourly precipitation in Norway}

\vspace{6mm}

{\large Anita Verpe Dyrrdal$^{1,2}$, Alex Lenkoski$^{3,*}$, Thordis L. Thorarinsdottir$^{3}$, Frode Stordal$^{2}$}

\vspace{6mm}

\noindent $^{1}$ The Norwegian Meteorological Institute, PO Box 43 Blindern, 0313 Oslo, Norway \\
$^{2}$ University of Oslo, Department of Geosciences, PO Box 1047 Blindern, 0316 Oslo, Norway\\
$^{3}$ Norwegian Computing Center, PO Box 114 Blindern, 0314 Oslo, Norway\\
$^{*}$ Corresponding author: {\tt alex@nr.no}
\end{center}

\vspace{10mm}

\begin{abstract}

\noindent Spatial maps of extreme precipitation are a critical component of flood estimation in hydrological modeling, as well as in the planning and design of important infrastructure. This is particularly relevant in countries such as Norway that have a high density of hydrological power generating facilities and are exposed to significant risk of infrastructure damage due to flooding. In this work, we estimate a spatially coherent map of the distribution of extreme hourly precipitation in Norway, in terms of return levels, by linking generalized extreme value (GEV) distributions with latent Gaussian fields in a Bayesian hierarchical model.  Generalized linear models on the parameters of the GEV distribution are able to incorporate location-specific geographic and meteorological information and thereby accommodate these effects on extreme precipitation.  Our model incorporates a Bayesian model averaging component that directly assesses model uncertainty in the effect of the proposed covariates.  Gaussian fields on the GEV parameters capture additional unexplained spatial heterogeneity and overcome the sparse grid on which observations are collected.  Our framework is able to appropriately characterize both the spatial variability of the distribution of extreme hourly precipitation in Norway, and the associated uncertainty in these estimates.    
\end{abstract}

\noindent\textbf{Keywords}: Generalized Extreme Value Distributions; Short-term extreme precipitation; Latent Gaussian processes; Return Levels; Uncertainty Assessment; Markov Chain Monte Carlo\\

\newpage

\section{Introduction}

Heavy rainfall over a short period of time often causes damage to infrastructure and thus represents an economic challenge as well as a threat to human safety. Such intense events are driven by complex spatio-temporal processes and are usually characterized by limited predictability and small spatial extent.  Estimation of the distribution of these events is exacerbated by a relatively sparse observational network. Nevertheless, in the planning and design of important infrastructure, such as roads and railways, dams, and urban environment, there is a great need for spatially continuous estimates of extreme short-duration precipitation. The need for meteorological information on smaller time-scales than a day is also becoming a requirement in hydrological modeling. In addition to the large spatial variability and relatively few observational sites, the complex terrain and different weather systems present in Norway further complicate such a task. 

Most official data and products from the Norwegian Meteorological Institute (MET Norway) are freely available for use, distribution and processing, see \url{http://met.no/English/Data_Policy_and_Data_Services/}.  This includes weather station data as well as gridded data products for daily temperature and precipitation at 3-hour temporal resolutions \citep{Tveitoetal2002, Mohr2009, Janssonetal2007, VormoorandSkaugen2013}.  The aim of the current study is to investigate the feasibility of producing gridded data sets of extreme hourly precipitation for Norway in terms of return levels based on hourly precipitation measurements from a relatively sparse network of observation stations combined with geographic and other meteorological information.  Here, the extremal properties of the available measurements are distributed in space through their relationship to the covariates which are collected on a considerably denser grid.


To accommodate both the diversity of precipitation patters present in Norway and account for the difficulty in data collection, we specify a hierarchical framework consisting of several components, which is estimated via Bayesian methods.  This involves specifying a generalized extreme value (GEV) distribution at each point in space.  The parameters of these GEV distributions then depend on location-specific variables, implying a structure similar to generalized linear modeling.  The complicated dynamics of extreme precipitation in Norway lead to heterogeneity in the manner that these variables affect the GEV parameters. To accommodate such overdispersion, a Gaussian field is used to allow for local adaptivity. Our strategy follows that of \cite{Davisonetal2012} who compare such a latent variable approach to methods based on copulas and max-stable random fields when applied to summer maximum daily rainfall in the Plateau region of Switzerland. Both \cite{Davisonetal2012} and \cite{ApputhuraiandStephenson2013} found that a latent variable approach is capable of computing the spatial distribution of marginal properties, which is our main objective.
 
In the model applied here we introduce Bayesian inference to make use of any prior knowledge and to obtain a measure of uncertainty, which has long been a shortcoming in return level estimation in Norway. Such a Bayesian Hierarchical Model (BHM) is estimated via Markov chain Monte Carlo (MCMC) methods, and our particular implementation is freely available in the {\tt R} package {\tt spatial.gev.bma}.  As discussed below, our algorithm is constructed such that little tuning is necessary on the part of the user, by relying of second-order Taylor series expansions to construct focused Metropolis-Hastings (M-H) proposals \citep{RueHeld2005}. While purely algorithmic in its innovation, such a development alleviates considerable burden when attempting to fit such highly structured models.  \cite{Cooleyetal2007} were the first to apply this type of model for daily precipitation threshold exceedance. They estimate parameters describing the Generalized Pareto distribution, and were able to produce maps of return levels for daily precipitation in Colorado, US. \cite{GaetanandGrigoletto2007} use a spatio-temporal BHM to assess trends in extreme rainfall over the Triveneto region (Italy) and \cite{SangandGelfand2009} apply a similar model to study extreme precipitation events from an interpolated dataset in the Cape Floristic Region of South Africa. \cite{GhoshandMallick2011} also propose a spatio-temporal BHM to model extreme precipitation events in the US, incorporating spatial and temporal information explicitly at the data level. \cite{CooleyandSain2010} and \cite{Schliepetal2010} study output from regional climate models via a spatial BHM, and \cite{ReichandShaby2013} propose a new BHM for analyzing max-stable processes, and apply this to analyze annual maximum precipitation using RCM output from the eastern US. 

Here, we apply a BHM to spatially interpolate the parameters of a GEV distribution for hourly precipitation in Norway, with the aim of producing return level maps. We believe this is currently among the best methods to create spatially continuous high-resolution maps that further include a measure of uncertainty. The maps can easily be updated and improved with increasing time series lengths and future observational sites. 

The rest of the article is organized as follows. Section~\ref{sec:data} introduces our data. We then introduce our BHM framework in Section~\ref{sec:modeling} our model fitting to the Norwegian data. Section~\ref{sec:results} presents some comparisons and results for the Norwegian data while Section~\ref{sec:conclude} contains some concluding discussion.  Much of the technical material related to fitting the BHM is supplied in the Appendix.

\section{Data}\label{sec:data}

\subsection{Hourly precipitation measurements}

\begin{figure}[!htbp]
\begin{center}
\includegraphics[width = 0.5\linewidth]{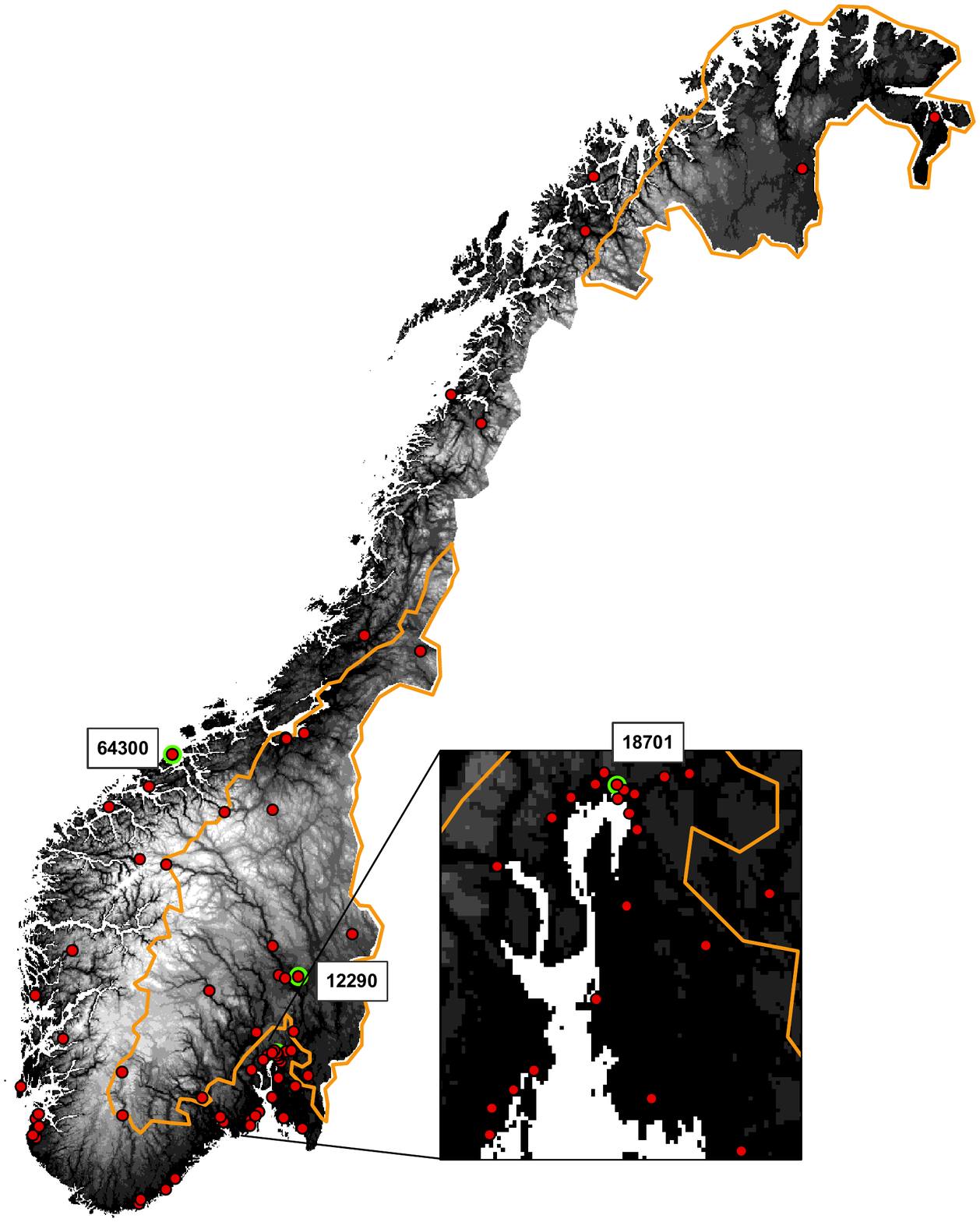} 
\end{center}
\caption[stations]{\label{fig:fig1}Map of Norway with observation stations indicated by red dots and the boundary of areas with dominated summer precipitation indicated with orange lines. The topography is shown in gray scale, with black denoting sea level and white denoting a height of approximately 2500m. The Oslofjord with Oslo located at the head of the fjord, is enlarged in the right square. Three stations analysed in subsequent sections are indicated by green circles.}
\end{figure}

Precipitation in Norway falls in three categories: frontal, orographic and convective. Most of the precipitation is frontal, caused by cyclone activity where warm and humid air in the south transitions with cold and dry air in the north. Orographic precipitation is caused by high speed vertical transmission of air, also called orographic lifting, observed in coastal mountain regions. Orographic and frontal precipitation dominate the climate along the western coast of the country which receives most of its precipitation in autumn and winter. The western coast receives the largest amounts of total annual precipitation while hourly precipitation levels might not be very high. Convective, or showery precipitation, on the other hand, occurs in unstable air given vertical currents and usually occurs in the heat of summer.  Finnmark in the north and {\O}stlandet in the south-east, see also Fig.~\ref{fig:fig1}, are somewhat sheltered from the large frontal systems which mainly come from the west and these regions are dominated by summer precipitation. While the total annual precipitation in these areas is relatively low, intense showers are common, particularly in the warmer south. There are important differences in the spatial structure of daily and hourly precipitation extremes in Norway. While daily extremes are higher in the Southwest where frontal precipitation dominates, hourly extremes are more closely associated with convective events which dominate the Southeast. For further information, see \url{http://met.no/English/Climate_in_Norway/}. 

Two types of rain gauges are used to measure hourly precipitation in Norway: Tipping bucket and weight pluviometer. The first tipping bucket stations were established in the spring of 1967 and the first weight pluviometer stations in December 1991. While some weight pluviometer stations are associated with technical difficulties resulting in erroneous values, the quality of the tipping bucket measurements is generally known to be good.  The data used in the current study have undergone a ``cleaning-process'' ({\em J.Mamen}, 2012, personal communication), removing unrealistic values and obvious errors.  Due to gaps in the data series caused by missing data and the removal of erroneous values, the data has been reduced to annual maxima. Our data set thus consists of the annual maxima from 59 tipping bucket stations and 10 weight pluviometer stations, which time series vary in length from 10 to 45 years.  In addition to the station network being sparse, the spatial distribution is highly inhomogeneous. As shown in Fig.~\ref{fig:fig1}, the majority of the stations are located in the south, especially in the surroundings of Oslo. This feature is, however, partly justified by the fact that the southern parts often experience the most intense and local showers, requiring a denser network. A lack of observations obviously introduces uncertainty and represents a challenge when attempting to distribute the statistical characteristics in space. 

\subsection{Gridded spatial covariates}\label{sec:datasets}

The explanatory variables (hereafter referred to as covariates) in our model, which serve to distribute the statistical characteristics of the extreme hourly precipitation in space are generated from gridded datasets on a $1 \times 1$ km$^2$ grid, covering the Norwegian mainland.  A list of the covariates we use is given in Table~\ref{tab:cov}.  Geographic information is obtained from a digital elevation model (DEM) based on a 100 m resolution terrain model from the Norwegian Mapping and Cadastre Authority (Kartverket) \citep{Mohr2009}.  Here, we consider latitude, longitude, elevation and distance to sea as potential geographic covariates.  

\begin{table}[!hbtp]
  \caption{Gridded spatial covariates included in the generalized linear models on the parameters of the GEV distribution.}\label{tab:cov}
  \centering
  \vspace{2mm}
  \begin{tabular}{lll}
    \toprule
    \textbf{Covariate} & \textbf{Abbreviation} & \textbf{Source}\\
    \midrule
    Latitude & \textsl{lat} & Digital elevation model \\
    Longitude & \textsl{lon} & Digital elevation model\\
    Elevation & \textsl{elev} & Digital elevation model\\
    Distance to sea & \textsl{distSea} & Digital elevation model\\
    Mean June-July-August temperature & \textsl{JJAtemp} & Daily temperature grid\\
    Mean annual precipitation & \textsl{MAP} & Daily precipitation grid\\
    Mean summer (April-October) precipitation & \textsl{MSP} & Daily precipitation grid\\
    Mean number of wet days & \textsl{wetDays} & Daily precipitation grid\\
    \bottomrule
  \end{tabular}
\end{table}

Additionally, we consider temperature and precipitation climatological covariates.  MET Norway produces a gridded dataset for daily temperature based on measurements at over 200 locations, interpolated to a map for the period 1957-today.  A residual kriging approach is applied for the interpolation, using terrain and geographic position to describe the deterministic component \citep{Tveitoetal2002, Mohr2009, Janssonetal2007}. We reduce this dataset to a single spatial covariate by taking the mean temperature during the months of June, July and August over all available years.  As discussed above, this climatological information may be related to the intensity of summer showery precipitation. 

MET Norway's gridded dataset for daily precipitation results from an interpolation of precipitation measurement at approximately 400 locations and it is also available for the period 1957-today \citep{Tveitoetal2002, Mohr2009, Janssonetal2007}. For the interpolation, triangulated irregular networks (TINs) were applied; a precipitation TIN based on measured precipitation and an elevation TIN based on the altitude at the meteorological stations.  Furthermore, a terrain adjustment was performed on the precipitation grid based on the assumption that precipitation increases by 10\% per 100 m up to 1000 m above sea level (masl) and 5\% above that \citep{Forland1979,Forland1984b}.  We extract three different spatial covariates from this dataset to capture the spatial variability in the climatological precipitation patterns: The mean annual precipitation, the mean summer (April-October) precipitation and the mean number of wet days per year.

While the geographic variables are considered relatively accurate, large uncertainties are associated with the climate datasets. For daily data these uncertainties are mainly related to the gridding procedure, particularly in regions with complex topography and a sparse network of stations, where the influence of a single station may cause biases.  The terrain adjustment on daily precipitation \citep{Engesetetal2004b, Saloranta2012} also adds additional uncertainty.  However, especially as we have performed smoothing in terms of temporal averages, we assume the spatial distribution to be sufficiently accurate for our purposes.  

\section{Methods}\label{sec:modeling}

\subsection{Extreme value statistics}

Extreme value theory provides a framework to model the tail of probability distributions.  Let $V_1, \ldots, V_n$ denote continuous, univariate random variables that are assumed to be independent and identically distributed.  If the normalized distribution of the maximum $\max\{ V_1, \ldots, V_n \}$ converges as $n \rightarrow \infty$, then it converges to a generalized extreme value (GEV) distribution \citep{FisherTippett1928}.  For this reason, the GEV distribution is commonly used to model block maxima such as the annual maxima.  Alternatively, if the full data series is available,  extreme value theory states that exceedances (the amounts that the observations exceed a given threshold $v$) should approximately follow a generalized Pareto (GP) distribution as $v$ becomes large and the sample size $n$ increases \citep{Pickands1975, Cooleyetal2007}.  \cite{Coles2001} provides an introduction to the statistical applications of extreme value theory. 

As our data are given by the annual maxima only, we employ the GEV modeling framework.   Let $\mathcal{S}$ denote the spatial region of interest (e.g. Norway) and $s\in\mathcal{S}$ a specific site in this region.  Our focus is on $y_{ts}$, the maximum hourly precipitation at location $s$ in a year $t$.  We assume that $y_{ts}$ follows a GEV distribution with spatially dependent parameters, 
\bl\[
y_{ts} \sim \text{GEV}(\mu_s,\sigma_s,\xi_s). 
\]\el
That is, the density of $y_{ts}$ is given by 
\bl\begin{equation}\label{eq:gev_density}
pr(y_{ts}| \mu_s, \kappa_s, \xi_s) = \kappa_s h(y_{ts})^{-(\xi_s + 1)/\xi_s} \exp \Big( - h(y_{ts})^{-\xi_s^{-1}}\Big), 
\end{equation}\el
for $h(y_{ts}) > 0$ with 
\bl\[
h(y_{ts}) = 1 + \xi_s \kappa_s (y_{ts} - \mu_s).
\]\el
The GEV distribution has three parameters which in our parameterization are location $\mu_{s} \in \real$, inverse scale $\kappa_{s} \in \real_+$ and shape $\xi_{s} \in \real$. The distribution is often parameterized using the scale $\sigma_s = 1/\kappa_s$ rather than the inverse scale \citep[e.g.][]{Coles2001}. However, the current parameterization is common in Bayesian contexts, for instance in the R-INLA toolbox \citep[\url{http://www.r-inla.org},][]{Rueetal2009}, and is chosen since derivations related to posterior densities are considerably easier in this representation.  

The tail behavior of the GEV distribution is driven by the value of the shape parameter $\xi_s$ and generally falls in three classes; the Fr\'echet type ($\xi_{s}>0$) has a heavy upper tail, the Gumbel type ($\xi_{s} \rightarrow 0$) is characterized by a light upper tail, and the Weibull type ($\xi_{s}<0$) is bounded from above.  The shape parameter thus provides vital information on the statistical properties of the variable of interest and is, concurrently, difficult to estimate due to the involved parametric form of the density in \eqref{eq:gev_density} as a function of $\xi_s$.  Note that the model formulation in \eqref{eq:gev_density} assumes stationarity in time.  While non-stationarity might generally be a more realistic assumption, for instance due to the effects of climate change, our data records are only 10 to 45 years.  This simultaneously renders the inclusion of non-stationarity assumptions difficult due to lack of data and reduces the risk of the data being severely affected by long-term non-stationarities.

The goal of the current analysis is to provide spatial measures of extreme hourly precipitation.  A common approach is to construct spatial maps of return levels.  The return level $z_p^{s}$ associated with the return period $1/p$ at location $s$ is the quantile that has probability $p$ of being exceeded in a particular year.  For the GEV density in \eqref{eq:gev_density}, it is given by 
\bl\begin{equation}\label{eq:retlev}
z_s^p = \mu_{s} -  (\kappa_s \xi_s )^{-1} \big[1 - \{ - \log(1 - p)\}^{-\xi_s}\big], 
\end{equation}\el
which is the quantile function of the GEV distribution function for the quantile $1-p$. 

\subsection{Modeling spatial dependence}\label{sec:gp}

The model in (\ref{eq:gev_density}) assumes that each location $s \in\mathcal{S}$ has its own set of parameters $(\mu_s, \kappa_s, \xi_s)$.  The spatial variability is the result of a number of factors related to the variation in terrain and climate.  To capture this information, we collect the additional covariates $\bs{x}_{s}$ listed in Table~\ref{tab:cov} which aim to incorporate these features.  The model for e.g. $\mu_s$ is then specified as
\bl\begin{equation}
\mu_s = \bs{x}^\top_s\bs{\theta}^{\mu} \label{eq:glm}, 
\end{equation}\el
and similarly for $\kappa_s$ and $\xi_s$. Here, we assume that $\bs{\theta}^{\mu} \in \bs{\Theta}^{\mu}_{M^\mu}$ for a fixed model $M^\mu$ in which some of the elements of $\bs{\theta}^{\mu}$ are assumed to take values on the real axis $\real$ while others may be restricted to be equal to zero. The constraint $\theta^\mu_i = 0$ implies that the $i$-th covariate does not influence the location parameter under the model $M^\mu$. Bayesian model averaging over all possible models is discussed in Section~\ref{sec:bma} below.   

The linear model in (\ref{eq:glm}) assumes that the variability in the GEV parameters $\mu_s$ is fully determined by the covariates $\bs{x}_s$.  In practice, there appears to be additional heterogeneity that is not directly captured by $\bs{x}_s$, requiring $\mu_s$ to be locally adaptive to overdispersion.  This is done by specifying the model
\bl\begin{equation}\label{eq:mu model}
\mu_s = \bs{x}^\top_s \bs{\theta}^\mu + \tau^{\mu}_s,
\end{equation}\el
where we follow \citet{Davisonetal2012} and give the overdispersion term $\tau^{\mu}_{s}$ a mean zero Gaussian Process prior with exponential decay, and hence any finite collection $(\tau_{s_1}^{\mu}, \dots, \tau_{s_T}^{\mu})$, with $s_t \in \mathcal{S}, t\in\{1,\dots,T\}$ is jointly normally distributed such that
\bl\begin{align}
E(\tau_{s_t}^{\mu}) &= 0 \label{eq:GPmean} \\
cov(\tau_{s_t}^{\mu}, \tau_{s_r}^{\mu}) &= \mc{K}_{\alpha^\mu,\lambda^\mu}(s_t,s_r) = \frac{1}{\alpha^{\mu}} \exp\left(-\frac{d_{s_ts_r}}{\lambda^{\mu}}\right),\quad s_t,s_r\in\mathcal{S}, \label{eq:GPcov}
\end{align}\el
where $d_{s_ts_r}$ is the Euclidean distance between locations $s_t$ and $s_r$.  The hyperparameters $\alpha^{\mu}$ and $\lambda^{\mu}$ determine the properties of this Gaussian process and we write this as $\tau_s^{\mu}\sim \mathcal{G}\mathcal{P}(\alpha^{\mu}, \lambda^{\mu})$.  

The models for $\xi_s$ and $\kappa_s$ are specified in a similar manner and our entire model may be written as
\bl\begin{align}\label{eq:full model}
y_{ts} &\sim \text{GEV}(\mu_s,\kappa_s,\xi_s) \nonumber \\
\mu_s &=  \bs{x}_s^\top \bs{\theta}^{\mu} + \tau^{\mu}_s\nonumber \\
\kappa_s &=  \bs{x}_s^\top \bs{\theta}^{\kappa} + \tau_s^{\kappa} \\
\xi_s &=  \bs{x}_s^\top \bs{\theta}^{\xi} + \tau_s^{\xi} \nonumber \\
\tau_s^{\nu} &\sim \mathcal{G}\mathcal{P}(\alpha^\nu,\lambda^\nu), \quad \nu \in \{ \mu, \kappa, \xi \}. \nonumber 
\end{align}\el 
The scale parameter $\sigma_s = 1/\kappa_s$ is often modeled with a logarithmic link function to ensure its positivity. We have chosen to use the identity link function for the inverse scale and to ensure $\kappa_s \in \real_+$, the proposal distribution is designed such that negative proposals are automatically rejected. In practice, we find that negative values are very rarely proposed once the chain is past the burn-in stage.  \cite{FriederichsThorarinsdottir2012} compare the logarithmic and the identity link functions for the scale parameter under frequentist inference in a prediction setting and find only a minor difference in the predictive performance, with the identity link resulting in minimally higher skill. 

This model imposes a conditional independence assumption on the full likelihood. Letting $\mathcal{Y}_o$ denote all observed responses, the likelihood satisfies
\bl $$
pr(\mathcal{Y}_o|\{\mu_s,\kappa_s,\xi_s\}_{s\in\mathcal{S}_o}) = \prod_{s\in\mathcal{S}_o}\prod_{t = 1}^{T_S} pr(y_{ts}|\mu_s,\kappa_s,\xi_s)
$$\el
which implies more generally that $y_{ts}$ and $y_{ts'}$ are conditionally independent for any $s\neq s'$ where $s,s' \in \mathcal{S}$, given the site-specific GEV parameters.  Such a conditional independence assumption is clearly a simplifying assumption, since neighboring sites would likely be affected by the same extreme events.  However, the purpose of the study is to construct characterizations of the marginal behavior at individual sites, for which this model is likely to give appropriate estimates.  We note that more involved methodologies \citep[see e.g.][among others]{SangandGelfand2009, ribatet_et_2012, GhoshandMallick2011} would be capable of incorporating such residual dependence.  The additional complexity imposed by these frameworks, and their demands on the data make them largely unhelpful for answering the marginal questions required in our data product. See Section~\ref{sec:full_results} for a further investigation of this feature.

Inference is performed via Markov chain Monte Carlo (MCMC) under the model in \eqref{eq:full model} where each component of the model is updated in turn.  The joint update of the regression parameters $\bs{\theta}^\nu$ and the model $M_\nu$ for $\nu \in \{ \mu, \kappa, \xi\}$ is discussed in the next section.  The updates for the Gaussian processes $\{ \tau^\nu_s\}_{s \in \mathcal{S}_o}$ and the corresponding hyperparameters $\alpha^{\nu}$ and $\lambda^{\nu}$ for $\nu \in \{ \mu, \kappa, \xi\}$ together with the associated prior assumptions are discussed in the Appendix.  Here, we use second-order Taylor expansions of the log-likelihood of the model in \eqref{eq:gev_density} to construct Gaussian proposal densities, thereby eliminating the need for user-defined tuning parameters for the proposal distributions, see e.g. Chapter 4.4 of \cite{RueHeld2005}.  

\subsection{Bayesian model averaging}\label{sec:bma}

We now discuss updating the regression parameters $\bs{\theta}^{\mu}$, $\bs{\theta}^\kappa$, $\bs{\theta}^{\xi}$ and their associated models $M^\mu$, $M^\kappa$, $M^\xi$.  The general strategy is the same for each of $\mu$, $\kappa$ and $\xi$.  For clarity of exposition, we thus discuss the updates in terms of a generic $\bs{\theta}$ and $M$ where we omit the parameter index.  Let $\mc{S}_o\subset \mc{S}$ denote the set of locations in which observations are collected, denote by $\bs{\Upsilon}$ the current vector of $\mu_s$, $\kappa_s$, or $\xi_s$ for $s \in \mc{S}_o$, that is $\bs{\Upsilon}_s =  \bs{x}_s^\top \bs{\theta} + \tau_s $, and let $\bs{X}_{\mc{S}_o}$ be the $|\mc{S}_o| \times |M|$ matrix of covariates, where $|\mc{S}_o|$ is the number of observation locations and $|M|$ is the number of regression parameters not restricted to zero under the model $M$.   Conditional on $\bs{\Upsilon}$, $\bs{X}_{\mc{S}_o}$ and the associated hyperparameters $\alpha$ and $\lambda$, the full conditional distribution of $\bs{\theta}$ and $M$ is independent of all other model parameters. That is, we aim to simultaneously update $\bs{\theta}$ and $M$ by sampling from the distribution
\bl\[
pr(\bs{\theta}, M| \bs{\Upsilon}, \bs{X}_{\mc{S}_o}, \alpha, \lambda) = pr(\bs{\theta}|M, \bs{\Upsilon}, \bs{X}_{\mc{S}_o}, \alpha, \lambda) pr(M|\bs{\Upsilon}, \bs{X}_{\mc{S}_o}, \alpha, \lambda)
\]\el
via a blocking, two step procedure.  First note that
\bl\[
pr(\bs{\theta}|M, \bs{\Upsilon}, \bs{X}_{\mc{S}_o}, \alpha, \lambda) \propto pr(\bs{\Upsilon}|\bs{\theta},\bs{X}_{\mc{S}_o}, \alpha, \lambda, M)pr(\bs{\theta}|M).
\]\el

We assign $\bs{\theta}$ a Gaussian prior distribution,  $\bs{\theta}| M \sim \mc{N}(\bs{\theta}_0, \bs{\Xi}_0)$, where $\bs{\Xi}_0$ is a matrix of dimension $|M| \times |M|$ and we have suppressed the zero elements of $\bs{\theta}$. It follows from the Gaussian process prior assumptions on $\tau_s$ that 
\bl\[
pr(\bs{\Upsilon}|\bs{\theta},\bs{X}_{\mc{S}_o}, \alpha, \lambda, M) = \mc{N}(\bs{X}^\top_{\mc{S}_o} \bs{\theta} , \mc{K}_{\alpha, \lambda}(\mc{S}_o,\mc{S}_o)).
\]\el
Standard results \citep[see e.g.][]{Hoff2009} then yield the posterior distribution 
\bl\begin{equation}\label{eq:theta posterior}
\bs{\theta}|M, \bs{\Upsilon}, \bs{X}_{\mc{S}_o}, \alpha, \lambda \sim  \mc{N}(\hat{\bs{\theta}}, \bs{\Xi}), 
\end{equation}\el
where
\bl\begin{align*}
\bs{\Xi} &= \bs{X}^\top_{\mc{S}_o} \mc{K}_{\alpha, \lambda}(\mc{S}_o,\mc{S}_o)^{-1} \bs{X}_{\mc{S}_o} + \bs{\Xi}_0\\
\hat{\bs{\theta}} &= \bs{\Xi}^{-1} \big[ \bs{X}^\top_{\mc{S}_o} \mc{K}_{\alpha,\lambda}(\mc{S}_o,\mc{S}_o)^{-1}\bs{\Upsilon} + \bs{\Xi}_0^{-1} \bs{\theta}_0 \big].
\end{align*}\el
The choice of the prior parameters $\bs{\theta}_0$ and $\bs{\Xi}_0$ is discussed in Section~\ref{sec:results} below.  

Sampling from the full conditional distribution of $M$ follows a conditional Bayes factor evaluation.  The full conditional distribution fulfills 
\bl\begin{align}
  pr(M|\bs{\Upsilon}, \bs{X}_{\mc{S}_o}, \alpha, \lambda) &\propto pr(\bs{\Upsilon}| \bs{X}_{\mc{S}_o}, \alpha, \lambda, M)pr(M) \nonumber \\
&= \Big[ \int_{\bs{\theta}} pr(\bs{\Upsilon}| \bs{\theta}, \bs{X}_{\mc{S}_o}, \alpha, \lambda, M)pr(\bs{\theta}|M)d\bs{\theta} \Big] \, \, pr(M). \label{eq:M integral}
\end{align}\el
We assume that all models with a non-zero constant term are {\em a priori} equally likely such that $pr(M) \propto 1$ while models with a zero constant term have {\em a priori} probability of zero. From \eqref{eq:theta posterior} it then follows that 
\bl $$
pr(\bs{\Upsilon}| \bs{\theta}, \bs{X}_{\mc{S}_o}, \alpha, \lambda, M)pr(\bs{\theta}|M) \propto (2\pi)^{-|M|/2}\exp\Big(-\frac{1}{2}\big[-2 \hat{\bs{\theta}}^\top \bs{\Xi}  \bs{\theta} + \bs{\theta}^\top \bs{\Xi} \bs{\theta}\big]\Big), 
$$\el
and we see that the integrand in \eqref{eq:M integral} is the canonical form of the kernel of a Gaussian distribution. Appropriate completion therefore gives
\bl $$
pr(M|\bs{\Upsilon}, \bs{X}_{\mc{S}_o}, \alpha, \lambda ) \propto |\bs{\Xi}|^{-1/2}\exp\Big(\frac{1}{2} \hat{\bs{\theta}}^\top \bs{\Xi}\hat{\bs{\theta}}\Big). 
$$ \el

For a joint update of $\bs{\theta}$ and $M$, we first sample a new model proposal $M'$ at random from the neighborhood of $M$. That is, one of the non-zero regression parameters in $M$ is set to zero or vice versa (excluding the constant term which is always included in the model).  The proposal $M'$ is then accepted with probability 
\bl\[
\min \Big\{ \frac{pr(M'|\bs{\Upsilon}, \bs{X}'_{\mc{S}_o}, \alpha, \lambda )}{pr(M|\bs{\Upsilon}, \bs{X}_{\mc{S}_o}, \alpha, \lambda )}, 1\Big\}.  
\]\el
In a second step, we sample a new value of the regression parameters  $\bs{\theta}$ according to \eqref{eq:theta posterior} based on the current model.  Given the posterior distributions, it is then simple to calculate the marginal posterior inclusion probability of a given covariate from the proportion of instances in which the corresponding regression parameter is non-zero.  However, it is generally not meaningful to consider the posterior probabilities of individual models when in the context of large hierarchies.\\
\indent The use of conditional Bayes factors and the MC3-within-Gibbs style of sampling above is just one approach to incorporating model uncertainty which has proven useful in a number of contexts involving model averaging in hierarchical models \citep[see][for related examples]{holmes_et_2002,karl_lenkoski_2011,cheng_lenkoski_2012}.  Other approaches, such as reversible jump MCMC \citep{green_1995}, spike-and-slab priors \citep{george_mcculloch_1993} or approximations involving information criteria \citep{raftery_1995} could also have been entertained.  In practice, we feel that each of these methods would have yielded comparable results.  We note, however, that our method involving CBFs offers the ability to completely integrate out the parameter set $\bs{\theta}$ when comparing two models (unlike reversible jump and spike-and-slab approaches) while still transitioning according to the exact posterior distribution (unlike approaches based on information criteria).  Further, in our study, mixing appeared straightforward, as shown below.

\subsection{Posterior return level maps}\label{sec:interpolation}

The MCMC algorithm returns a chain of length $R$ (after an appropriate burn-in period has been removed) with values for all parameters in the model above that approximate their joint posterior distribution. That is, in the $r$th iteration of the MCMC we have the elements
\bl\[
\big\{\bs{\theta}^\mu, \{\tau_s^{\mu}\}_{s\in\mc{S}_o}, \alpha^\mu, \lambda^{\mu}\big\}^{[r]},
\]\el
that fully describe the model for $\mu_s$, and a similar set for $\kappa_s$ and $\xi_s$.  We note that from these chains alone the posterior of the return level $z^p_s$ for any $s\in\mc{S}_o$ may be derived by calculating the return level in \eqref{eq:retlev},  
\bl\[
(z_s^p)^{[r]} = \mu_{s}^{[r]} -  (\kappa_s^{[r]} \xi_s^{[r]} )^{-1} \big[1 - \{ - \log(1 - p)\}^{-\xi_s^{[r]}}\big], 
\]\el
for $r = 1,\ldots,R$.  The sample $(z_s^p)^{[1]}, \dots, (z^p_s)^{[R]}$ then gives an MCMC approximation of the posterior distribution of $z^p_s$.

For locations $q\in\mc{S} \setminus \mc{S}_o$ we utilize the Gaussian process prior and the states of $\{\tau^\nu_{s}\}_{s\in\mc{S}_o}$ for $\nu \in \{ \mu, \kappa, \xi \}$ to interpolate  the relevant $\mu_q^{[r]}$, $\kappa_q^{[r]}$ and $\xi_q^{[r]}$ parameters at each stage $r$ of the MCMC output to the location $q$.  Suppose $\mc{A}$ and  $\mc{B}$ are two finite subsets of $\mc{S}$ and let $\mc{K}_{\alpha,\lambda}(\mc{A},\mc{B})$ be the $|\mc{A}|\times |\mc{B}|$ matrix with $[\mc{K}_{\alpha,\lambda}(\mc{A},\mc{B})]_{ab} = \mc{K}_{\alpha,\lambda}(a,b)$ for $a\in\mc{A}$ and  $b\in\mc{B}$, where we have omitted the parameters $\nu$ and $r$ for clarity.  Then, if $\tau_s \sim \mc{G}\mc{P}(\alpha,\lambda)$ it holds that
\bl\begin{equation}
\tau_q|\{\tau_{s}\}_{s\in\mc{S}_o} \sim \mc{N}(\hat{\tau}_q, \hat{\kappa}_q)\label{eq:impute}
\end{equation}\el
where 
\bl\begin{align*}
\hat{\tau_q} &= \mc{K}_{\alpha,\lambda}(q,\mc{S}_o)\mc{K}_{\alpha,\lambda}(\mc{S}_o,\mc{S}_o)^{-1}\bs{\tau}_{\mc{S}_0}\\
\hat{\kappa}_q &= \alpha - \mc{K}_{\alpha,\lambda}(q,\mc{S}_o)\mc{K}_{\alpha,\lambda}(\mc{S}_o,\mc{S}_o)^{-1}\mc{K}_{\alpha,\lambda}(\mc{S}_o,q),
\end{align*}\el
with $\bs{\tau}_{\mc{S}_o}$ the vector of current $\tau_s$ for $s\in\mc{S}_o$.  Thus, at iteration $r$ in the MCMC chain, we may estimate $(\tau_q^{\nu})^{[r]}$ using $\big\{\{\tau_s^{\nu}\}_{s\in\mc{S}_o}, \alpha^\nu, \lambda^{\nu}\big\}^{[r]}$ for $\nu \in \{ \mu, \kappa, \xi\}$ according to (\ref{eq:impute}) and obtain $\{ \mu_q, \kappa_q, \xi_q\}^{[r]}$, thereby deriving $(z^p_{q})^{[r]}$ and approximating the marginal posterior distribution for $z^p_q$ at all sites $q \in \mc{S}$. We note that this could be done in a joint manner (for all $\mc{S}$) by modifying (\ref{eq:impute}).  However, in practice such a joint estimation incurs a prohibitive computational overhead (particularly due to memory constraints) and is largely unnecessary, since the site-specific marginal distribution is of primary interest in constructing return level maps.
\section{Results}\label{sec:results}
This section shows some results from using the methodology above to estimate the spatial GEV distribution using our data from Norway.  The first subsection conducts a leave-one-out cross validation study and compares our full approach utilizing bayesian model averaging with several other options.   In the second subsection we investigate these results in-depth for three stations chosen in particular.  Finally, we conclude with a discussion of the fit of the model using all stations and highlight both the performance of our algorithm, as well as the ability to draw return level maps.
\subsection{Cross-validation}\label{sec:cv}
\indent We begin with a leave-one-out cross-validation (CV) study in which we compare our overall approach (which we call BMA for short) to three alternatives.  The first alternative includes all covariates and makes no attempt to model average and we refer to this approach as Full.  The second alternative (referred to as NoCovar) represents the other extreme: only the constant term, latitude and longitude are included, which is meant to investigate the additional benefit of the other covariates in the model fit.  Finally, we consider a case in which the shape parameter $\xi_s$ is set to a fixed value (referred to as Fixed).  

Estimation of the shape parameter $\xi$ is known to be extremely uncertain, particularly when time series are short, which is the case in this study. The value of $\xi$ for daily precipitation has been analyzed in several papers. \cite{PapalexiouandKoutsoyiannis2013} studied annual maximum daily rainfall of 15 137 records from all over the world, and declared the Fr\'echet distribution ($\xi>0$) to be ``the winner''. This distribution represents the lowest risk for engineering structures. \cite{Koutsoyiannis2004b} indicated a shape value of 0.15 as appropriate for daily precipitation in mid-latitude areas of the Northern Hemisphere after using several different methods of estimation. \cite{WilsonandToumi2005} fitted a GEV distribution to long daily precipitation records from the UK and found a mean $\xi$ estimate of 0.10. Daily precipitation in Norway was studied by \cite{Dyrrdaletal2014} who found that $\xi$ varies spatially according to dominating weather systems. Positive values are seen in the continental inland, while more negative values are seen along the coast in the south and west. $\xi$ for hourly precipitation is not well studied, due to fewer observations combined with larger spatial variance, however it is likely to be higher than for daily precipitation. This is confirmed in e.g. \cite{Overeemetal2010} and \cite{VandeVyver2012}. As there are substantial differences between the spatial distribution of daily and hourly precipitation in Norway, it is not feasible to transfer the spatial variability of $\xi$ for daily precipitation directly to the hourly precipitation. Instead, in the Fixed approach we choose to fix $\xi$ at a constant value of 0.15 over the entire country. This value is equivalent to the upper range of established values for daily precipitation.

\indent For each modeling framework, the model is run $69$ times, where in each instance one station is left out.  For reasons discussed in detail in Section~\ref{sec:full_results}, each instance is run for 200 000 iterations and the first 20 000 iterations are discarded as burn-in.  A single run takes approximately 2 hours on a 2.8 gHz multicore server using the present {\tt R} implementation.  In Section~\ref{sec:conclude} we discuss implementation issues that should prove to reduce this computing time dramatically.

\indent Now, suppose that site $s\in\mc{S}_0$ is the site that is left out.  The predictive distribution $pr(Y_s|\bs{Y}_{-s})$ is thus formed as discussed in Section~\ref{sec:interpolation} and this distribution is compared to the observations $\bs{Y}_s = \{Y_{1s},\dots,Y_{T_ss}\}$.  We consider two scores: The continuous rank probability score (CRPS) as well as the logarithmic score (LS).  Table~\ref{tab:cv_scores} shows the scores averaged over the $69$ sites, revealing two interesting features.  First, we note that CRPS and LS scores are often, in magnitude, unlikely to show substantial numerical differences.  However, while 2.525 and e.g. 2.520 are not far from each other numerically, such a difference is substantive. In Table~\ref{tab:cv_scores} we see that the BMA approach has slightly better CRPS and LS scores than the Fixed approach, both of which considerably outperform the Full and NoCovar approaches.  This indicates that the BMA approach is able to trade-off the desire for parsimony with the ability to chose covariates that have predictive impact.  This, in turn, yields a more realistic predictive distribution that is able to capture the full spread of the observations.  This leads to lower CRPS and LS scores than, most notably, the Full and NoCovar approaches.  Since the Fixed approach will introduce less uncertainty in the predictive distributions the increase in sharpness is rewarded.\\

\begin{table}\caption{Predictive Scores for the leave-one-out cross validation study}\label{tab:cv_scores}
\begin{center}
\begin{tabular}{lcc}
\hline\hline
 & CRPS & LS\\
\hline
BMA & 2.520 & 2.823\\
Full & 2.543 & 2.840\\
NoCovar & 2.542 & 2.839\\
Fixed & 2.525 & 2.826\\
\hline\hline
\end{tabular}
\end{center}
\end{table}

\begin{figure}
\begin{center}
\subfigure[Site 15720]{\includegraphics[width=.47\linewidth]{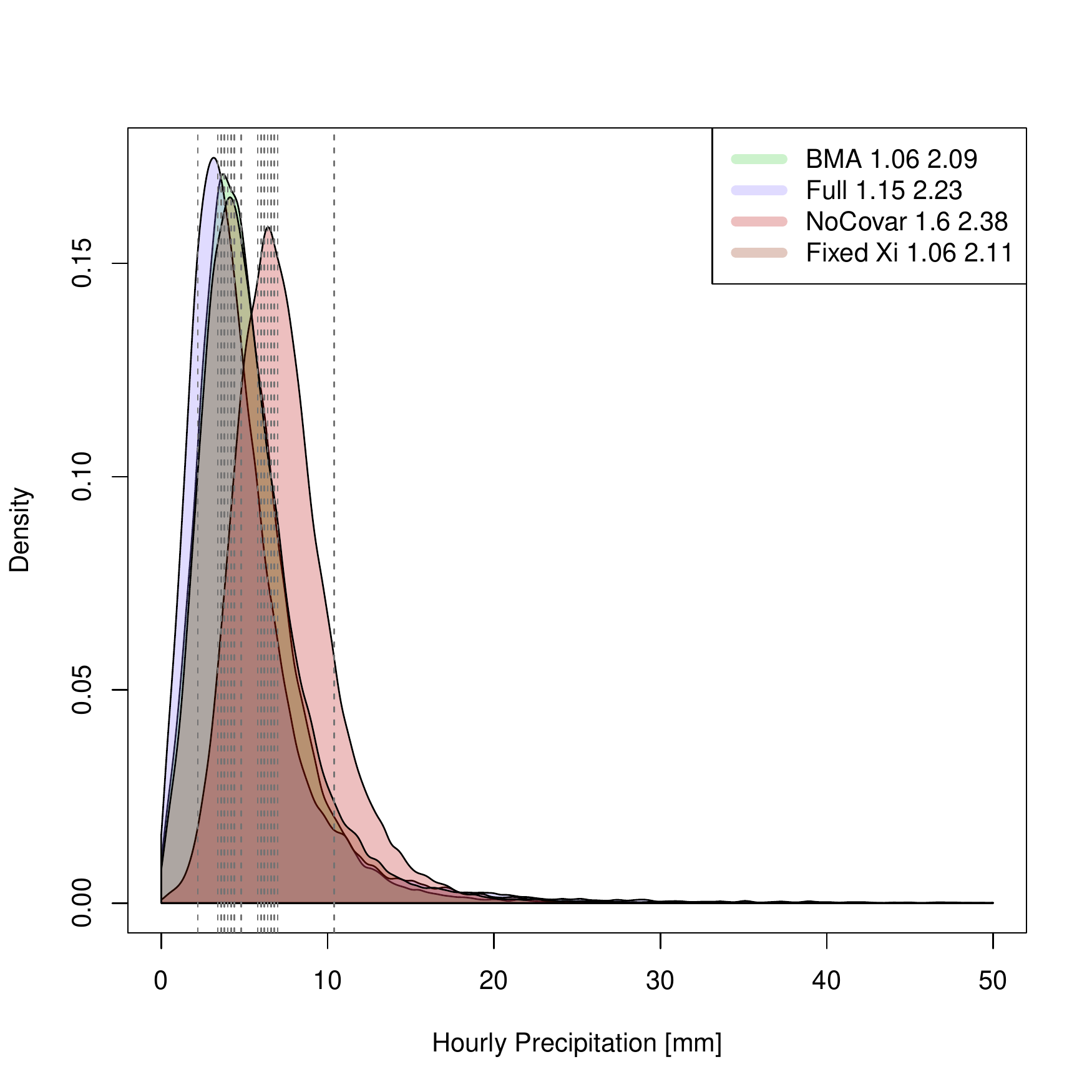}}
\subfigure[Site 40880]{\includegraphics[width=.47\linewidth]{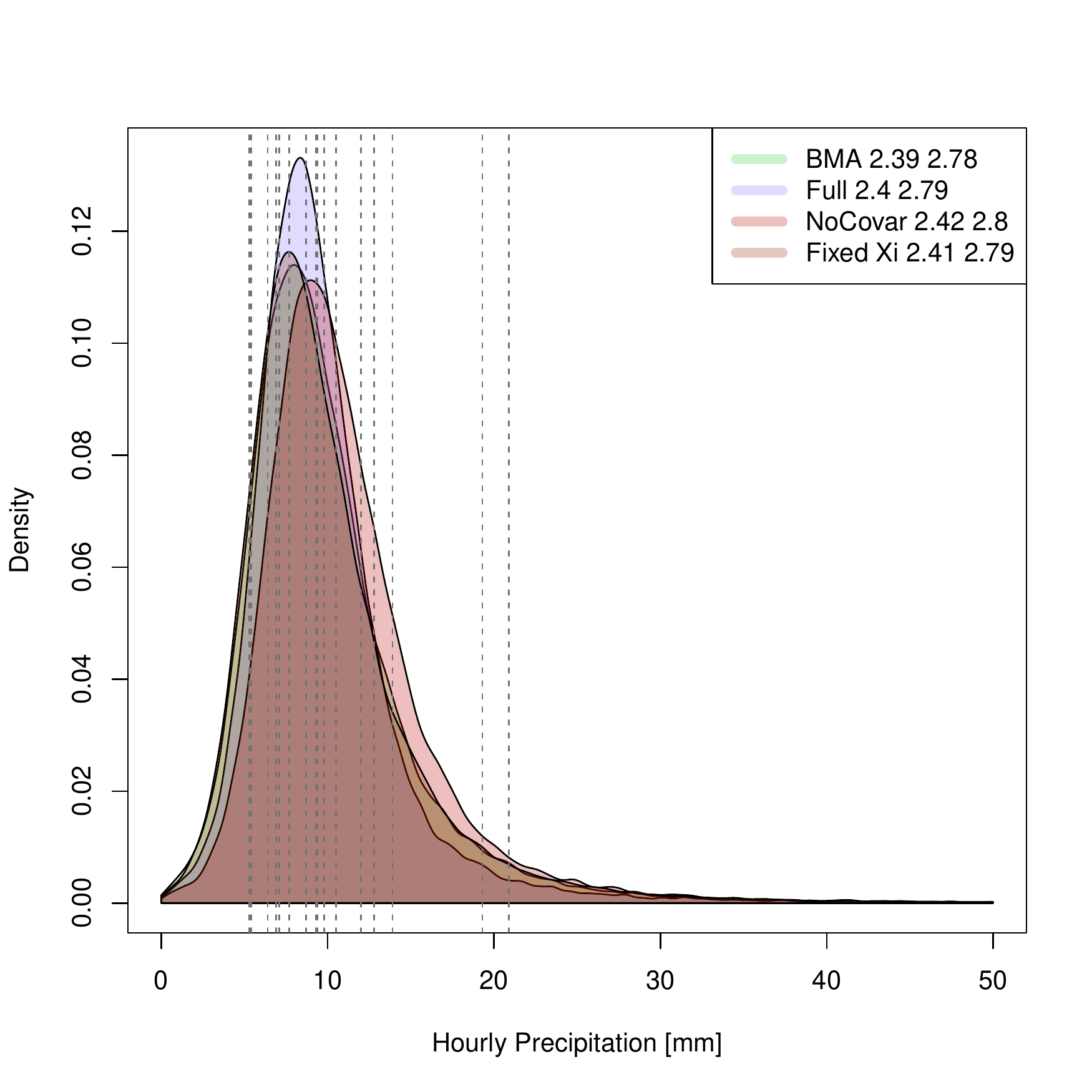}}
\end{center}
\caption{Out of sample predictive distributions for stations 15720 and 40880 in the cross validation study.  The dotted gray lines show the observed levels and the two numbers in the legend correspond to the CRPS and LS for each method.}\label{fig:cv_examples}
\end{figure}

\indent Figure~\ref{fig:cv_examples} shows the predictive distributions for Station 15720 (left panel) and Station 40880 (right panel), as examples of the relationship between predictive distributions in the various methods.  In these panels we see two features.  First, the predictive distribution involving no covariates beyond latitude and longitude tends to be centered somewhat differently than the other three methods (this is particularly pronounced for Station 15720).  Secondly, the method that does not model average tends to yield a more peaked predictive distribution, evidence of ``over-fitting'' which is shown clearly in the distribution for Station 40880.  The method that fixes $\xi$ and the BMA approach often have roughly similar predictive distributions.  These examples serve to therefore show that covariate information is indeed important, yet having the flexibility to remove covariates more appropriately reflects uncertainty in the predictive distribution.  Furthermore, the introduction of uncertainty in the shape parameter $\xi$ does not appear to markedly affect the bulk of the predictive distribution.

\clearpage
\subsection{Example of three stations}
\begin{figure}
\begin{center}
\subfigure{\includegraphics[width=.32\textwidth]{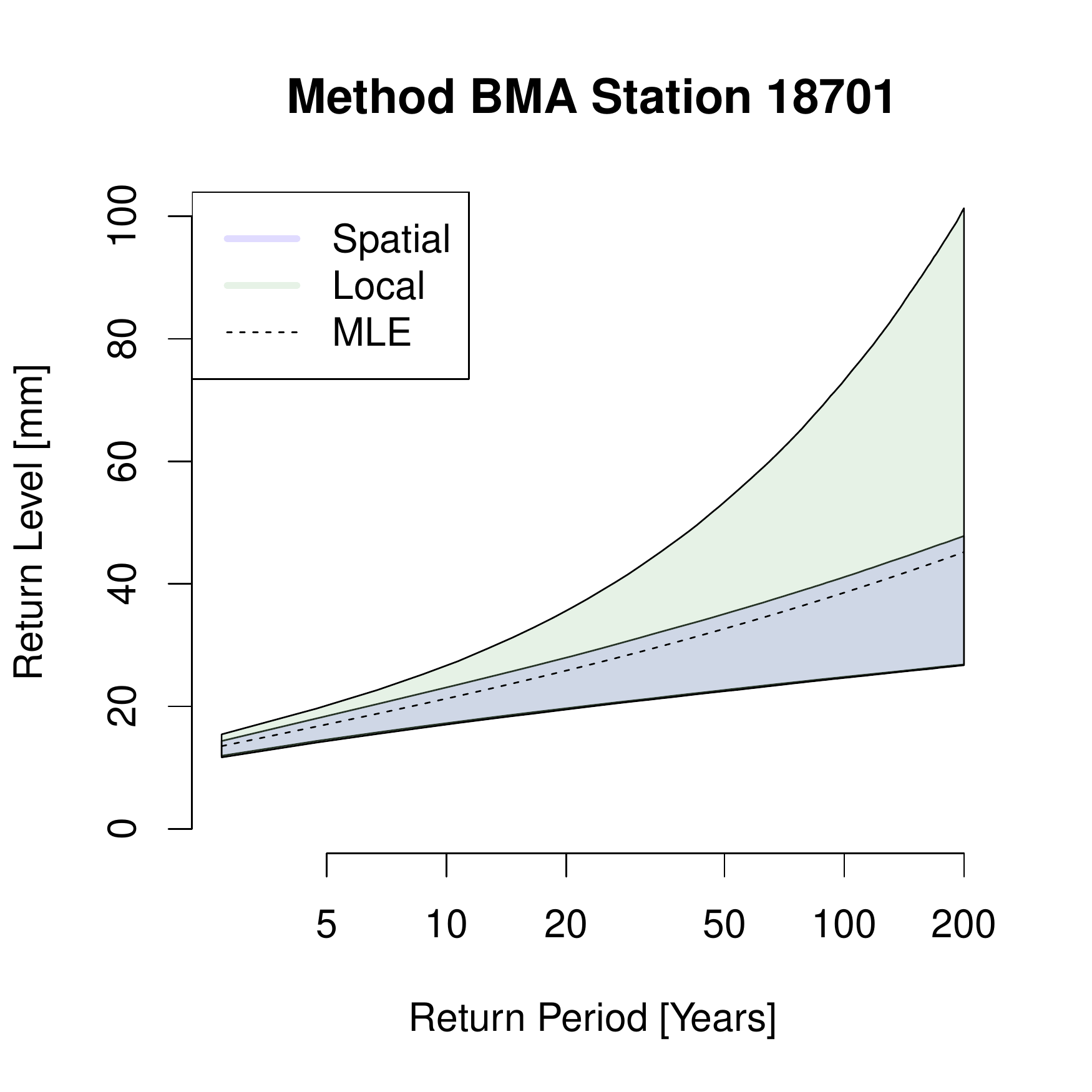}}
\subfigure{\includegraphics[width=.32\textwidth]{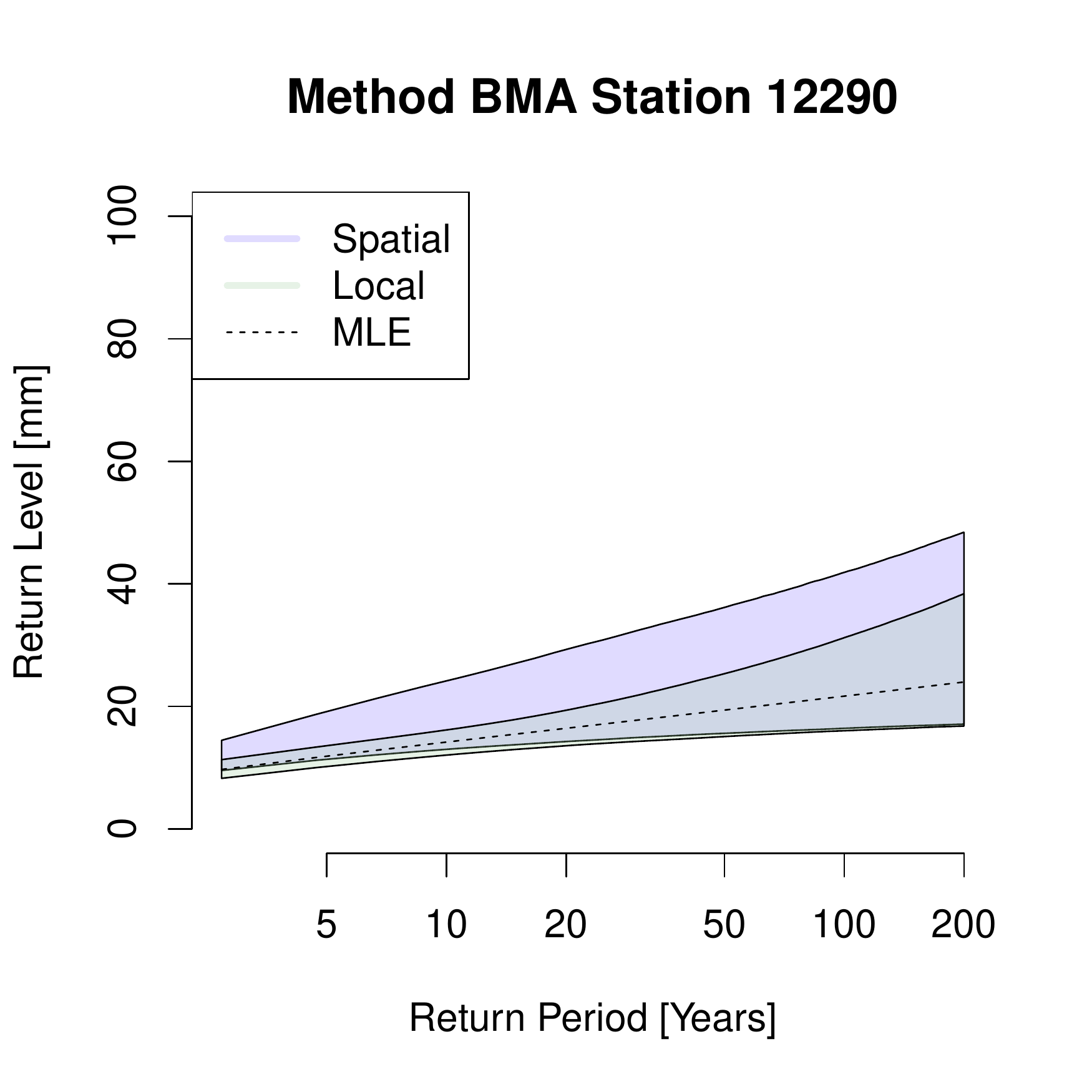}}
\subfigure{\includegraphics[width=.32\textwidth]{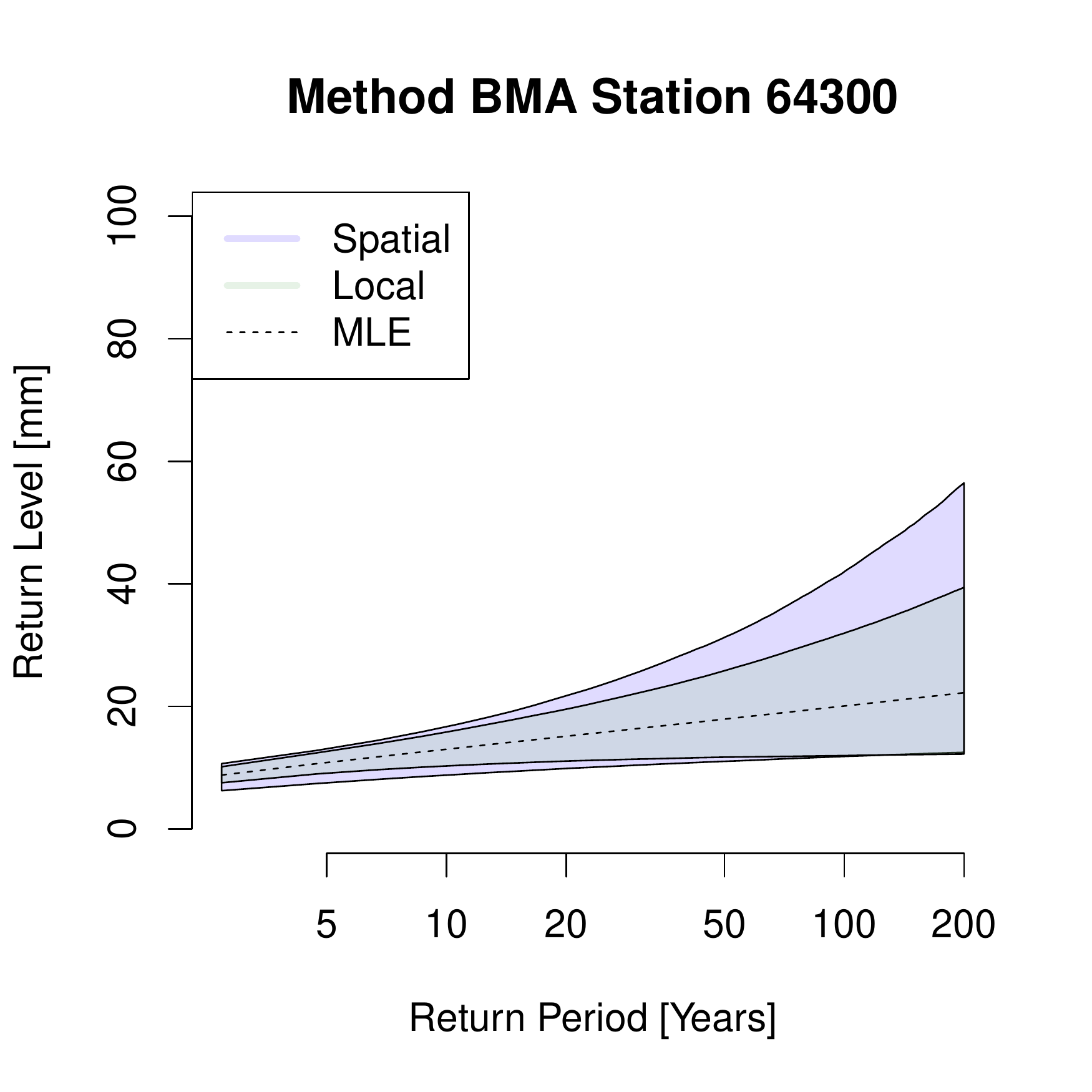}}
\subfigure{\includegraphics[width=.32\textwidth]{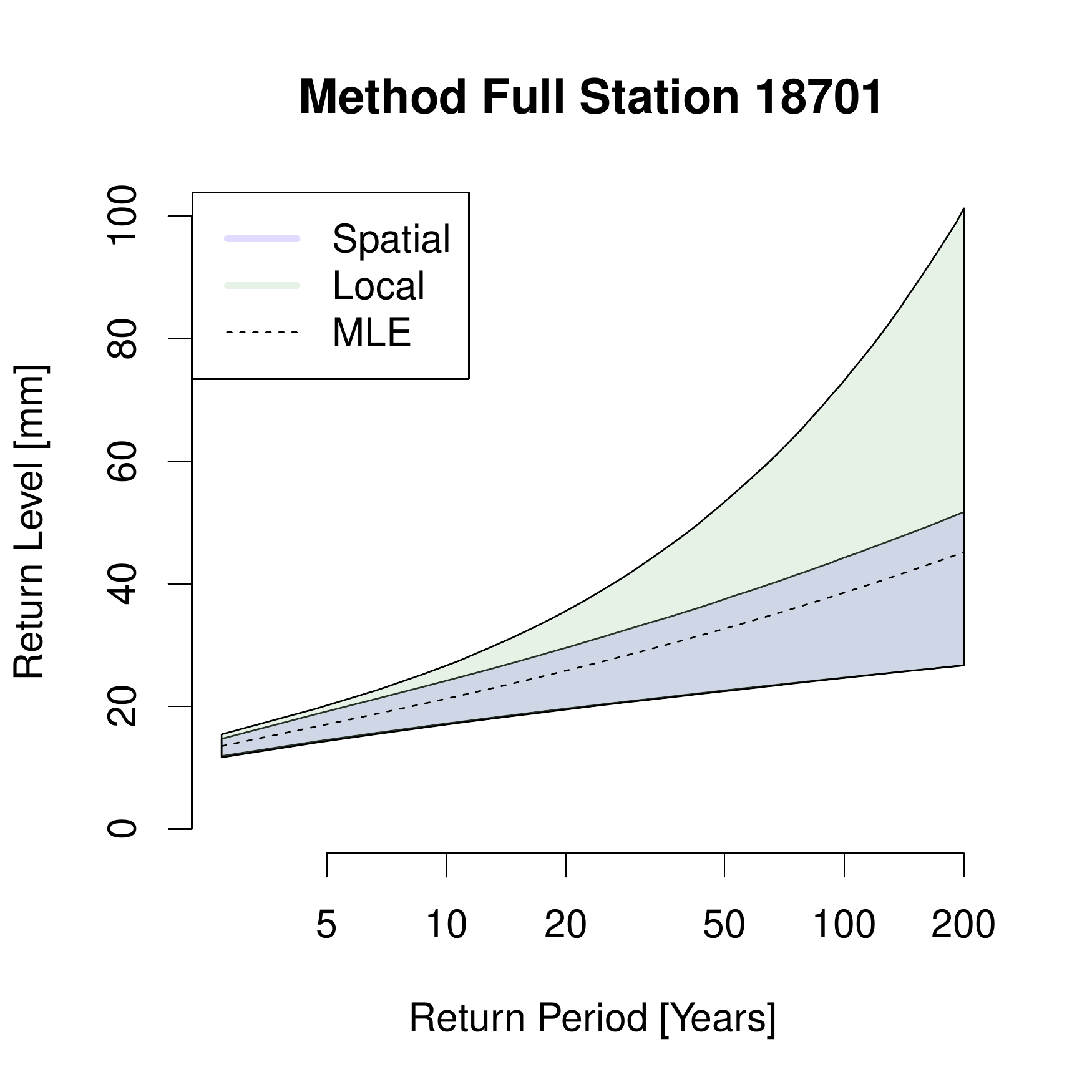}}
\subfigure{\includegraphics[width=.32\textwidth]{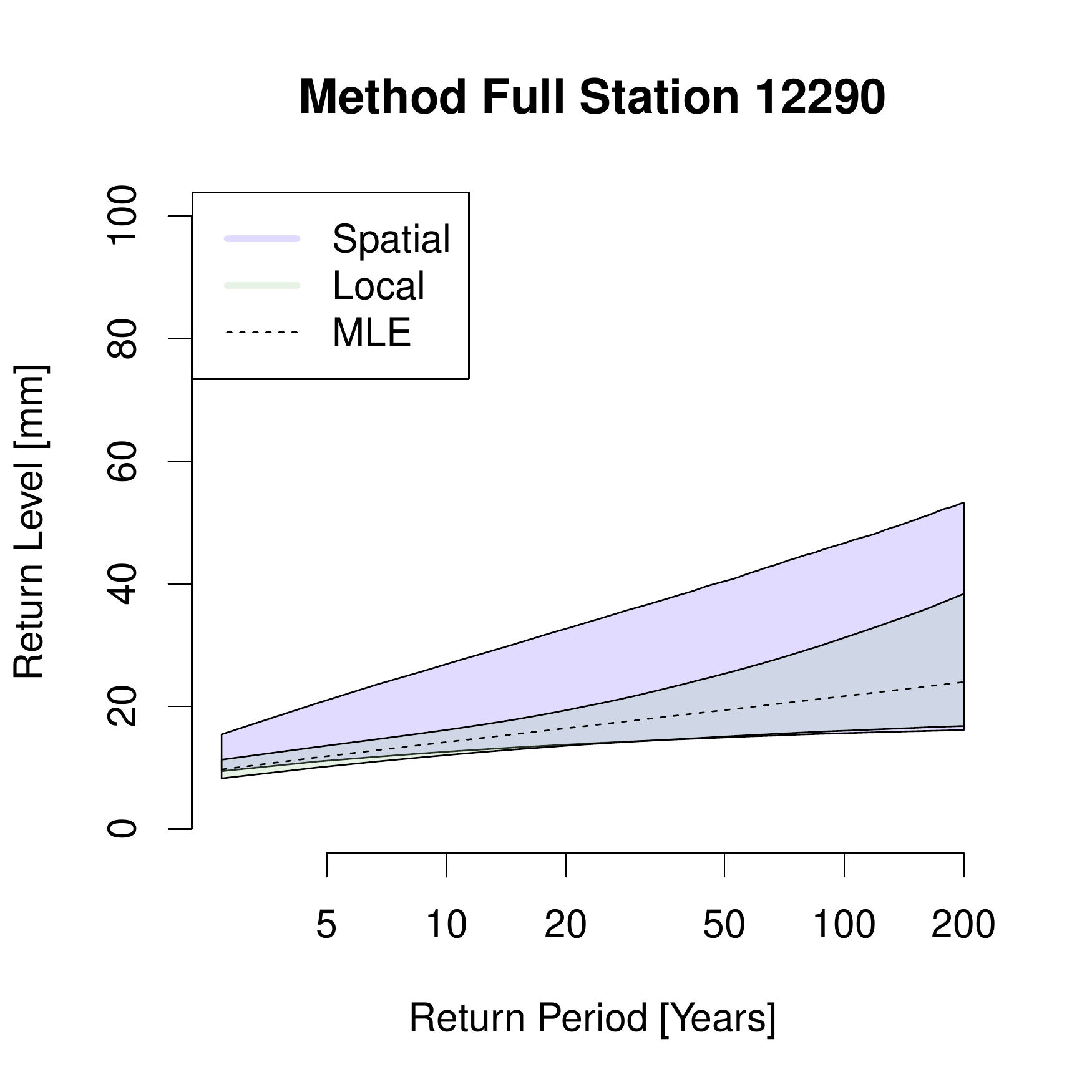}}
\subfigure{\includegraphics[width=.32\textwidth]{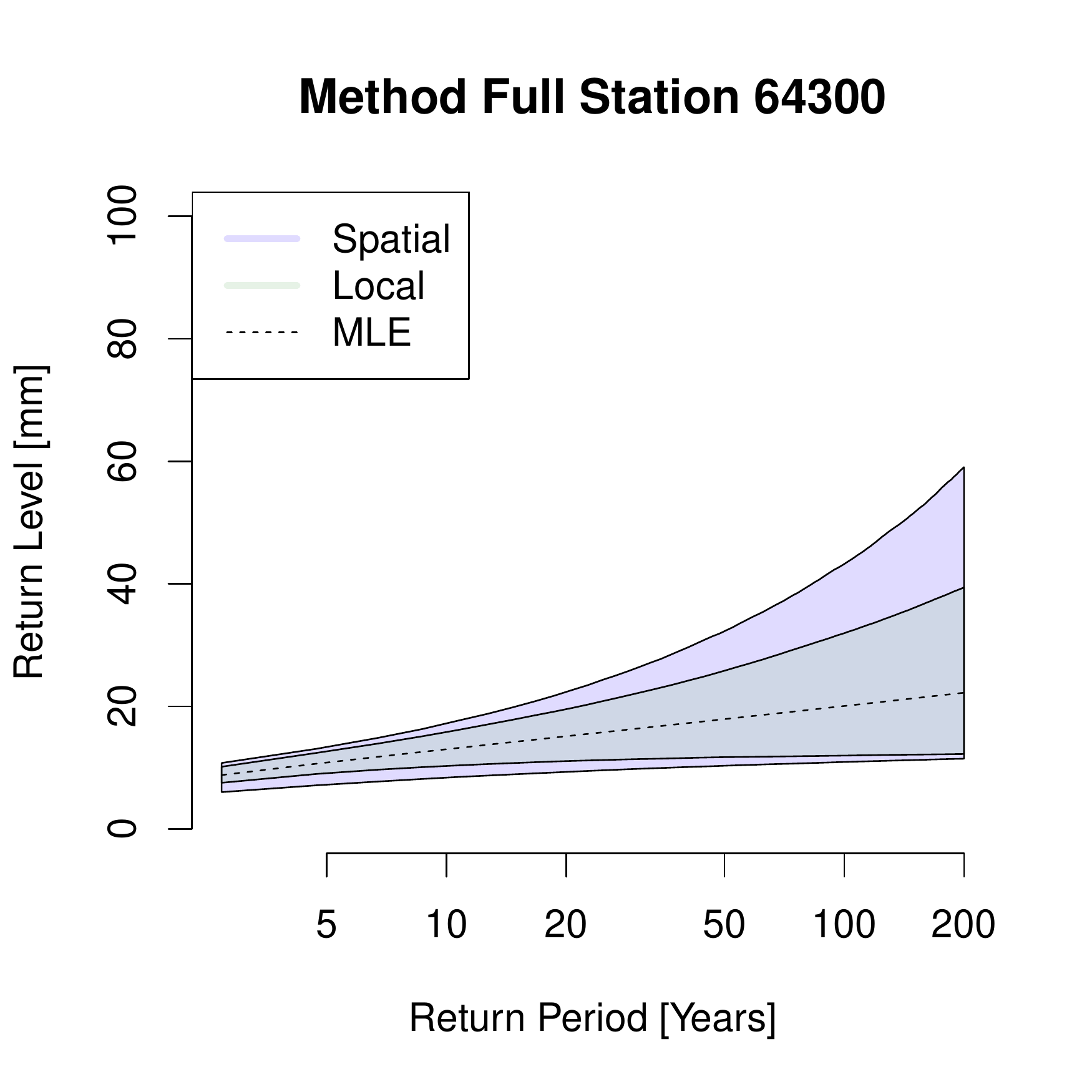}}
\subfigure{\includegraphics[width=.32\textwidth]{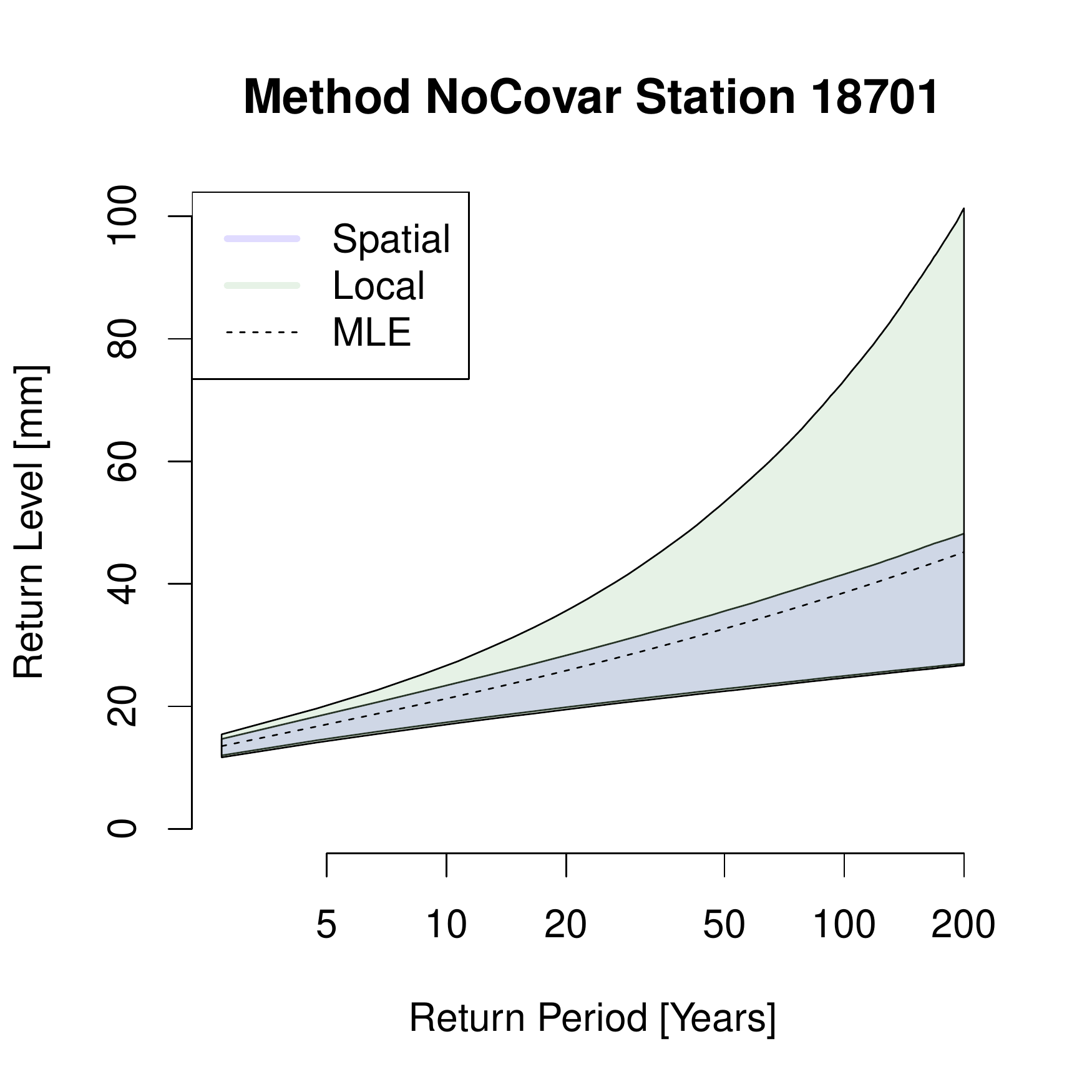}}
\subfigure{\includegraphics[width=.32\textwidth]{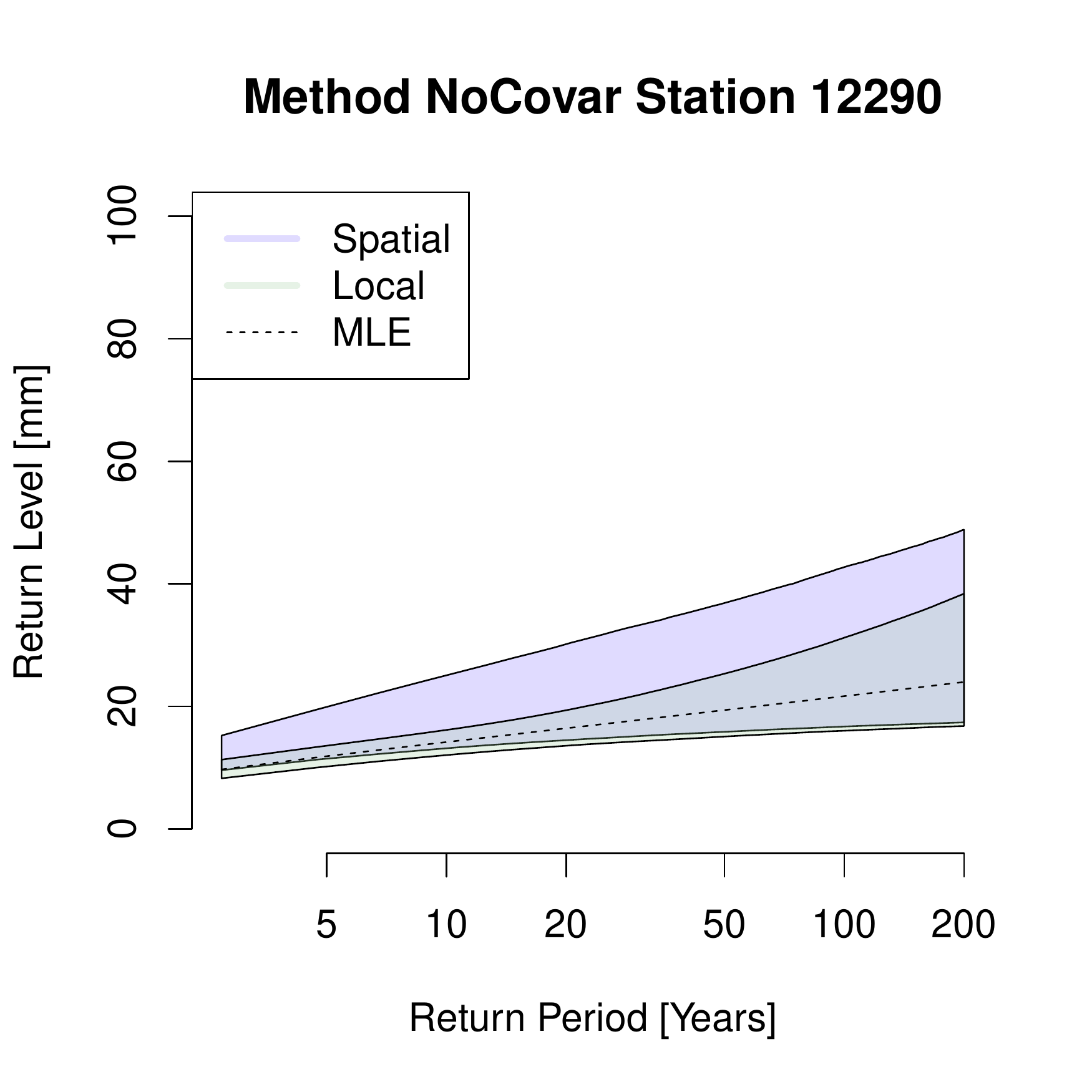}}
\subfigure{\includegraphics[width=.32\textwidth]{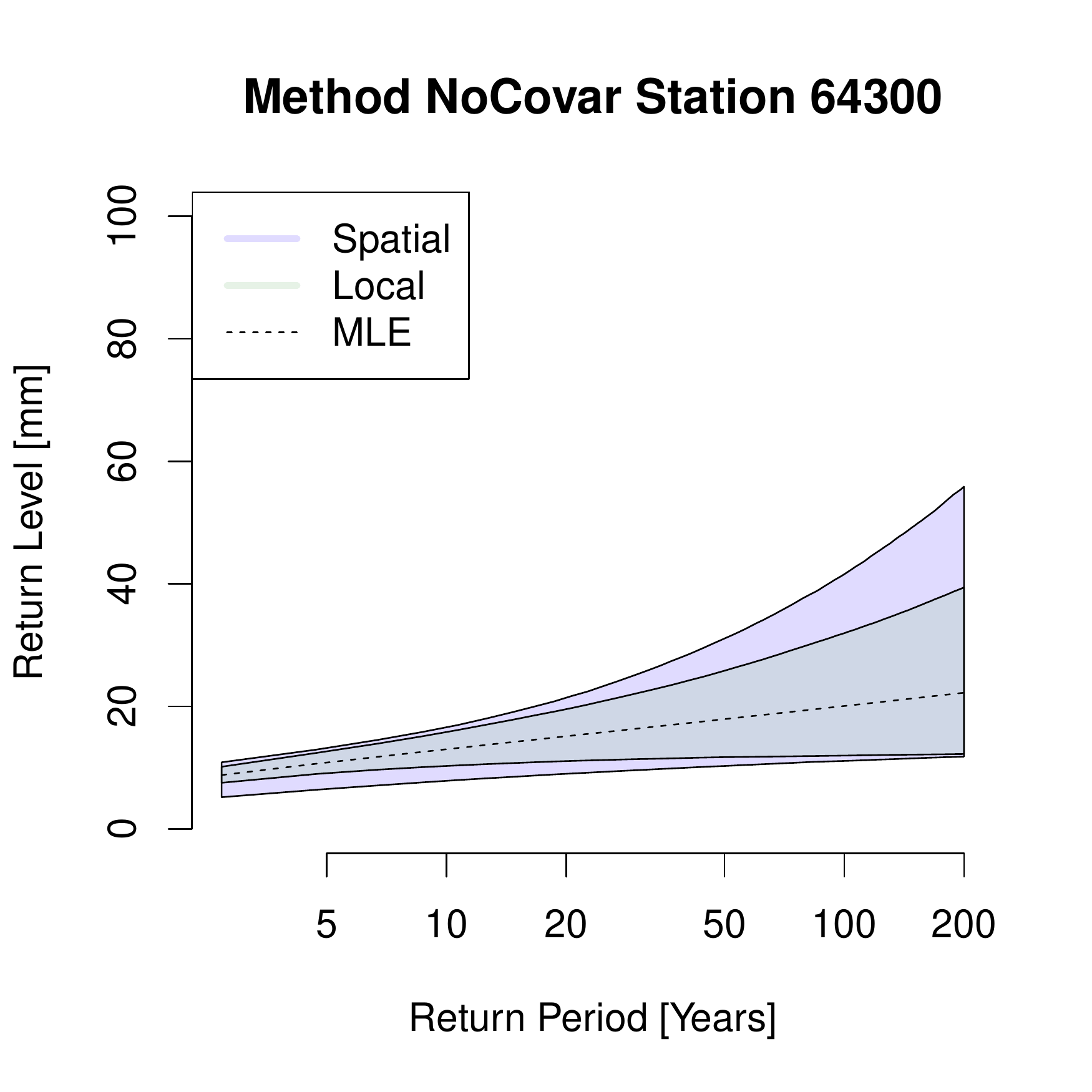}}
\subfigure{\includegraphics[width=.32\textwidth]{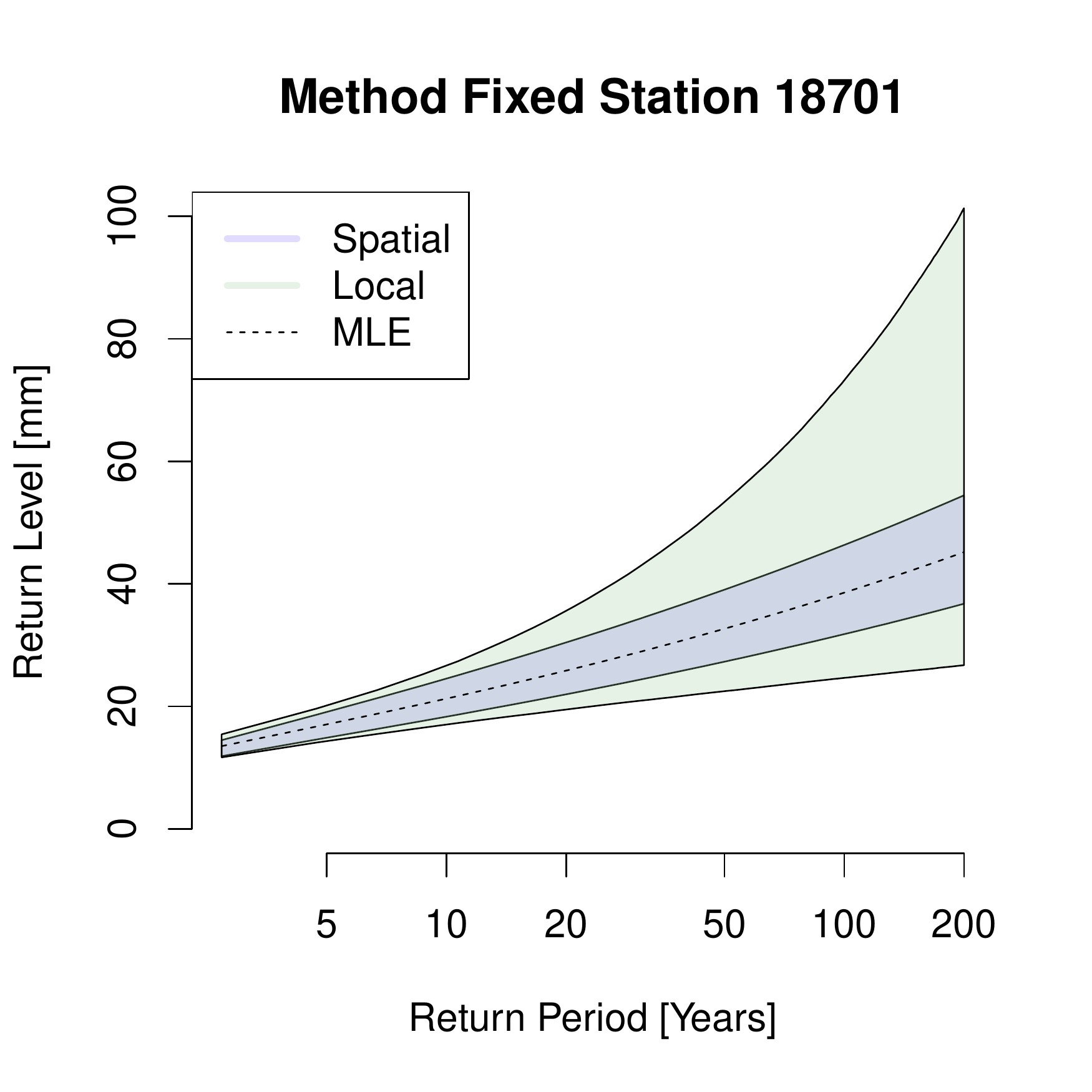}}
\subfigure{\includegraphics[width=.32\textwidth]{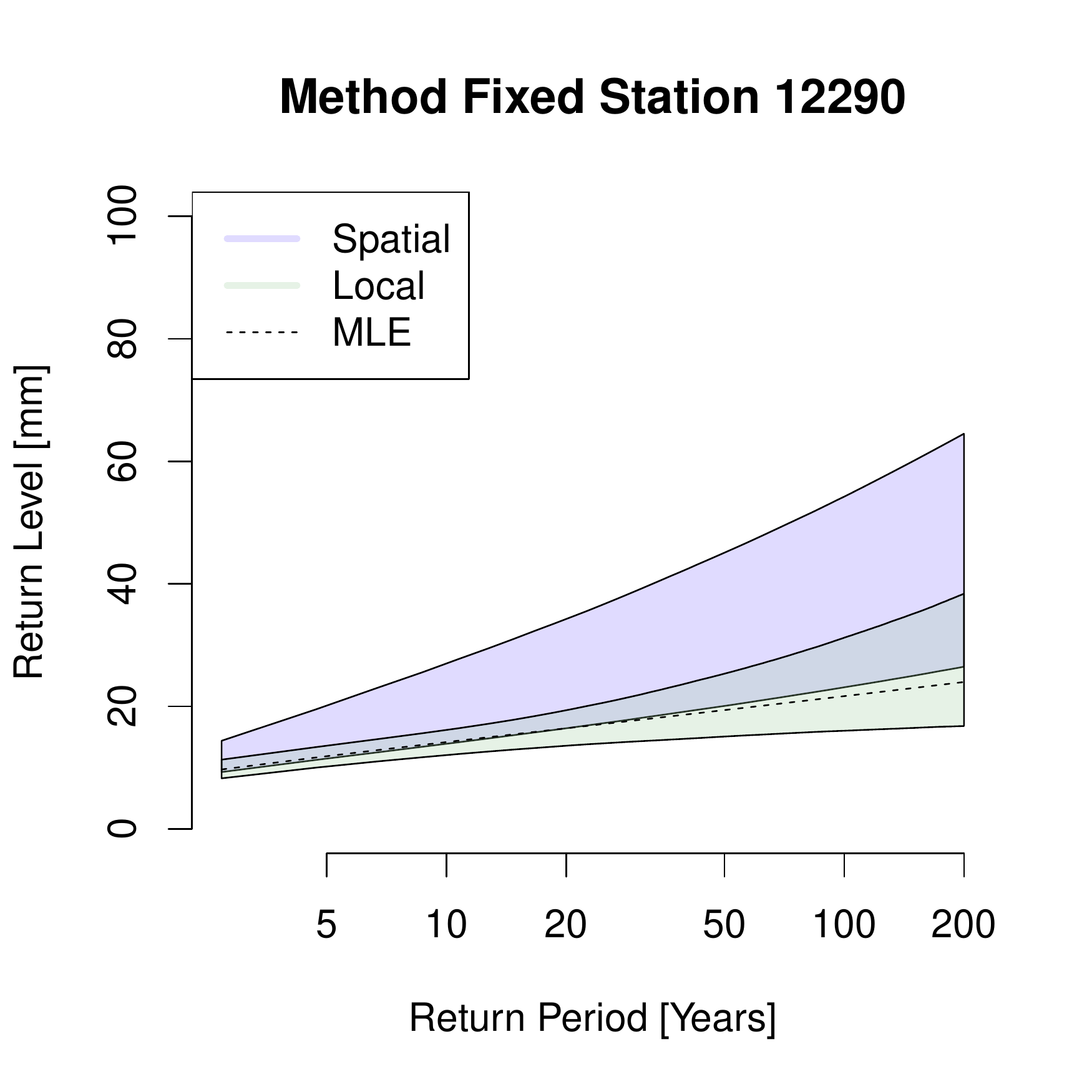}}
\subfigure{\includegraphics[width=.32\textwidth]{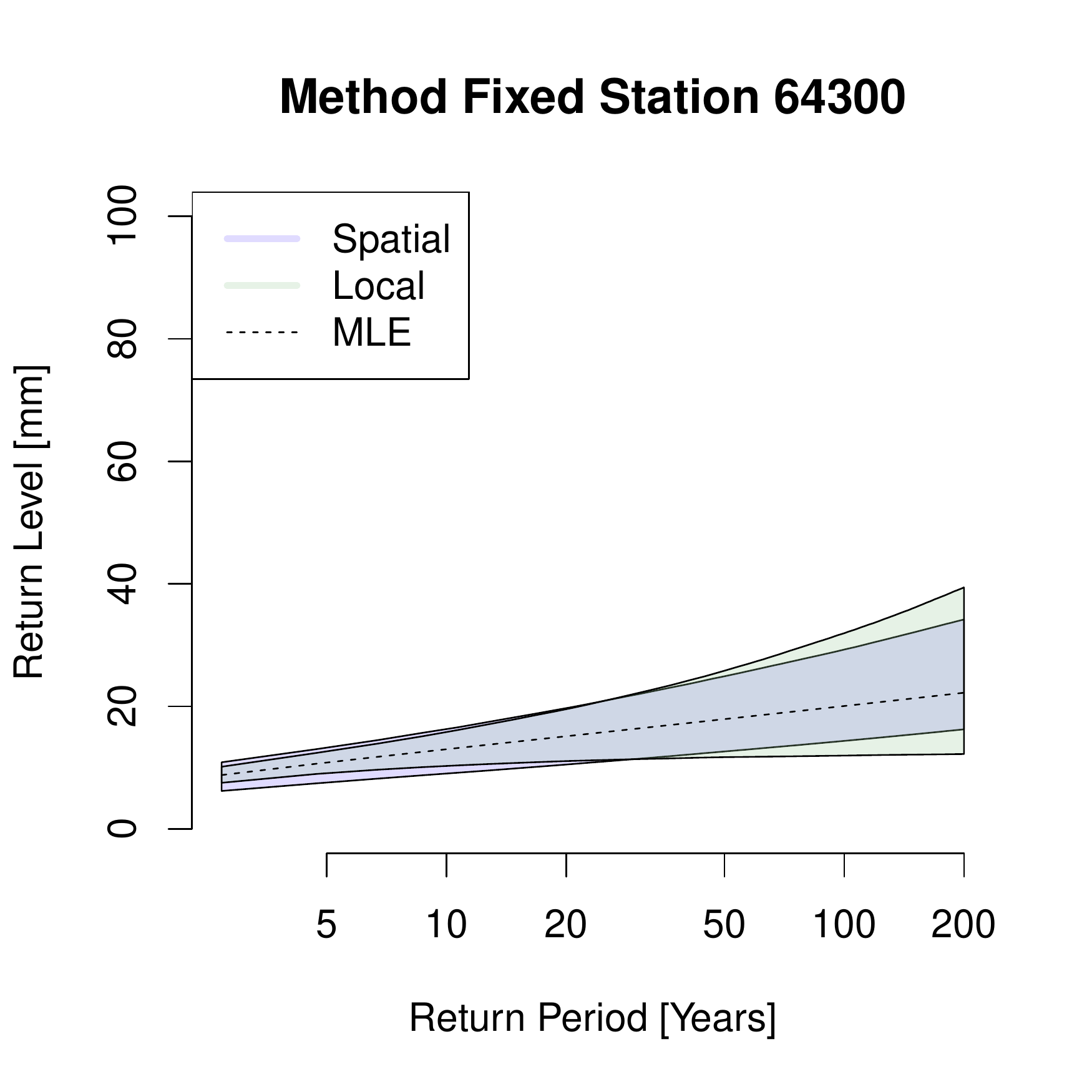}}
\end{center}
\caption{Return levels and observed returns for each method/station compared to a local bootstrapped MLE.}\label{fig:returns}
\end{figure}

Figure~\ref{fig:returns} shows the return levels from a leave-one-out CV study for three representative stations 18701 (Oslo, 94 masl), 12290 (Hamar, 132 masl) and 64300 (Kristiansund, 39 masl) (cf. Figure~\ref{fig:fig1}).  In each panel the return levels (and associated uncertainty at the 90\% level) from each spatial method (BMA, Full, NoCovar and Fixed) is compared to the MLE estimated locally at the site, with the local confidence bands given by bootstrapping. The first column shows the fit of the methods for the station at the head of the Oslo fjord, revealing several things.  First, there is across-the-board a reduction in the uncertainty in return levels from the spatial model versus the bootstrapped local version.  In our mind, this is sensible, since there is such a high concentration of observation sites in the vicinity of this location.  Secondly, we see that all methods contain the MLE return levels, with the Fixed method apparently performing best.  This is unsurprising, as the MLE for $\xi$ at this location is $.17$, and thus fixing $\xi$ to a value nearby will concentrate estimates of the return levels about those of the local estimate.

The second column of Figure~\ref{fig:returns} shows the estimated return levels for station 12290, located in the continental Southeast.  In this case, we see that the three methods that statistically estimate $\xi$ match the locally estimated return levels better than the method which fixes $\xi$.  This is because the local MLE for $\xi$ at this site is $.025$, much below $.15$.  Indeed, we see that return levels for this site are over-estimated as a result of fixing $\xi$ to too high a level.

Finally, the third column of Figure~\ref{fig:returns} shows results for the observation station on the west coast of Norway.  In this column we see that all methods contain the MLE return levels in their posterior predictive intervals.  For the three methods that estimate all parameters in the model, we see that the uncertainty is larger using the spatial model than a local bootstrapped MLE.  This seems sensible to us, as the network of observation sites is much more sparse here and thus the uncertainty around the return levels would be expected to be greater than when using the information at that site to estimate the model locally.  However, we note that the Fixed approach exhibits considerably less uncertainty in the estimated return levels.  We do not view this as a positive feature, as we believe that this is under-representing the true uncertainty at this site.

The general conclusion is that the spatial model provides an interpolation into sites without observations that is broadly consistent with return levels that would be estimated via maximum likelihood at the site when data are present.  This provides some evidence that the spatial modeling framework can yield useful interpolation over Norway.  Further, there is some evidence that incorporating the uncertainty in the shape parameter does not cause undue difficulty, more appropriately captures the spatial variability in this term and gives a more accurate characterization of return level uncertainty.

Taken together, the results of this section and Section~\ref{sec:cv} indicate the benefit of using the BMA framework. We have shown that BMA, along with Fixed, outperform the two other approaches in estimating the main characteristics of hourly precipitation (cf. Table~\ref{tab:cv_scores} and Figure~\ref{fig:cv_examples}). Further, when estimating extreme values, the three methods that allow $\xi$ to vary appear to offer estimates that are more consistent across Norway (cf. Figure~\ref{fig:returns}). The fact that the BMA approach performs well in both estimating the ''bulk`` of the predictive distribution and its tail, demonstrates the enhanced flexibility.

\subsection{Return level maps}\label{sec:full_results}
\begin{figure}
\begin{center}
\includegraphics[width = .45\linewidth]{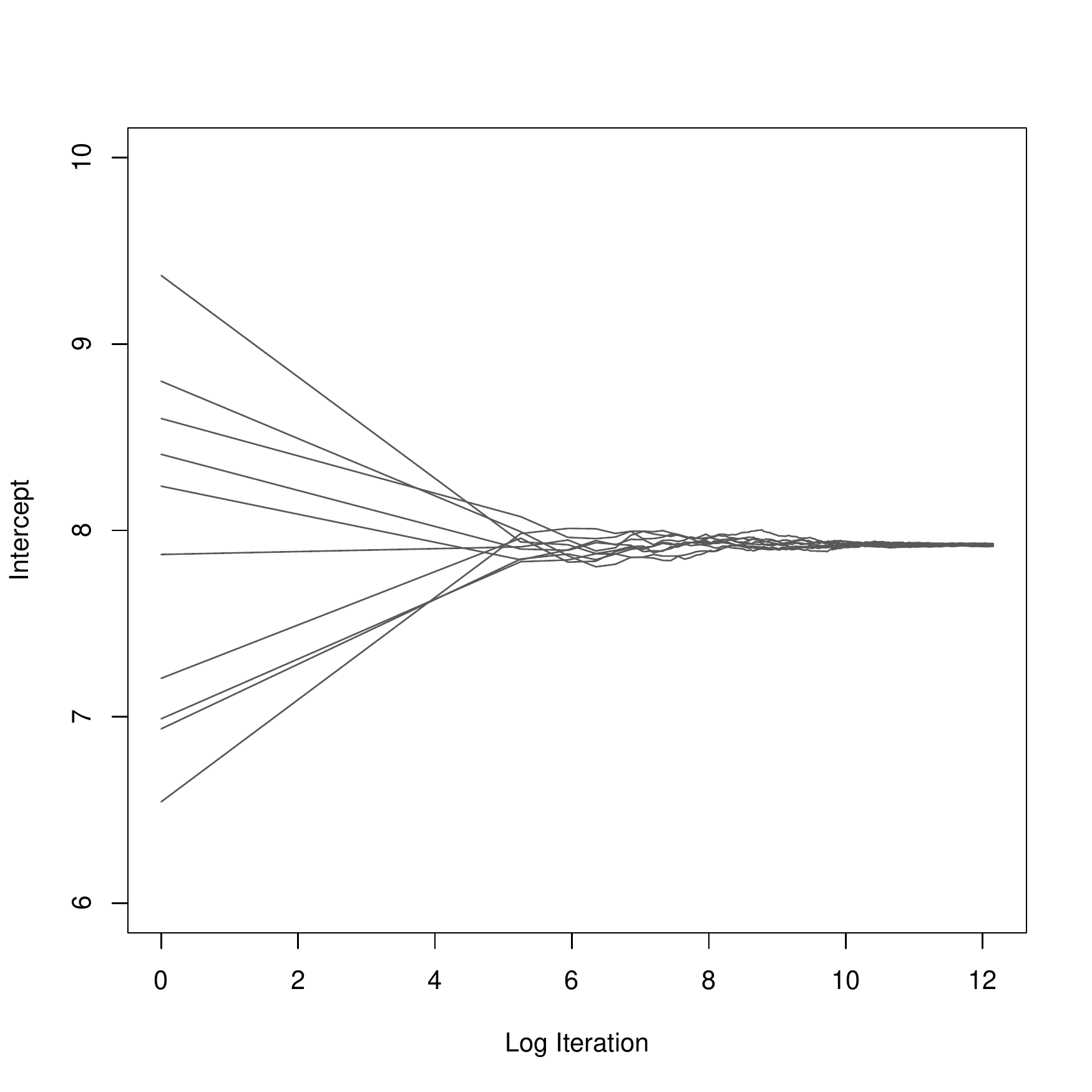}
\end{center}
\caption{Convergence assessment for the run over the full dataset.  We see that 15 separate chains run independently with different starting seeds all agree on the value of the intercept term after 180 000 iterations (with 20 000 iterations first used as burn-in).}\label{fig:convergence}
\end{figure}

Here we discuss the return level maps constructed using the full dataset.  As mentioned above, we ran all studies for 200 000 iterations and discarded the first 20 000 iterations as burn in.  Figure~\ref{fig:convergence} indicates why this chain length seemed appropriate.  In this figure we see the running estimate of the posterior mean on the intercept term of $\theta_\mu$, plotted by log iteration for 15 different chains run independently with different random seeds.  As shown in the figure, after about 100 000 iteration the 15 chains essentially agree on the value of the posterior expectation for the intercept and after 180 000 iterations these estimates are identical.  Other quantities considered show a similar agreement and imply that the chain length is sufficient for approximating the posterior distribution.  Appendix B details our choice of prior settings and further investigates the sensitivity of return levels maps to these settings.

Table~\ref{tab:linear_results} shows the estimates for linear terms taken from this run.  We see several interesting features from this table.  First, in the location term $\mu$, the MSP has the highest inclusion probability at 0.9, while JJAtemp, lat, lon, MAP and elevation also feature strongly, and all covariates have a non-negligible inclusion probability. The estimated regression coefficients for MSP and JJAtemp reveal strictly positive 95\% confidence bands. The combination of covariates with high inclusion probabilities accounts for both geographic (lat, lon, elev) and meteorological (MSP, JJAtemp, MAP) features that are known to influence short-duration precipitation in Norway. Summer indices seem essential as a majority of the most extreme hourly precipitation events occur during summer, and many events are a result of convective instability created by surface heating. Latitude and longitude are good covariate candidates due to the strong gradients in both temperature (north-south) and precipitation (east-west), although the orientation of the country from southwest to northeast represents a challenge. 
Elevation is likely to be important in regions where orographic precipitation plays a role.

The situation is much different for both the precision term $\kappa$ and shape $\xi$.  In both instances, there is considerable statistical uncertainty and no covariate is given appreciable inclusion probability. The most influential covariates, however, seems to be lat and lon. We note that the lower quantile of the constant on the precision term is, unintuitively, negative.  However, this is a result of the fact that the mean of the random effects $\tau_s^\kappa$ for $s\in\mc{S}_o$ is not forced to be 0 in our implementation. See Appendix B for a discussion of this aspect.

Table~\ref{tab:linear_results} also corroborates the values for $\xi$ that have been suggested in the literature, with a posterior mean of $.11$.  However, as can be shown from the wide band about this value (with a .025 quantile of -.65 and .975 quantile of .87), there is significant statistical uncertainty regarding this quantity and none of the presently collected covariates appear to have a substantive impact on these estimates. 
\begin{table}
\caption{Posterior estimates for the linear terms in the BHM under the BMA approach.  This table shows the probability that a given covariate is included in the model}\label{tab:linear_results}
\begin{center}
\begin{tabular}{l|c c c c|c c c c|c c c c}
\hline\hline
& \multicolumn{4}{c|}{Location ($\mu$)} & \multicolumn{4}{c|}{Precision ($\kappa$)} & \multicolumn{4}{c}{Shape ($\xi$)}\\
& Prob & Mean &2.5\% & 97.5\%& Prob & Mean &2.5\% & 97.5\%& Prob & Mean &2.5\% & 97.5\%\\
\hline
Intercept & 1 & 7.92 & 6.64 & 9.17 & 1 & 0.3 & -0.46 & 1.05 & 1 & 0.11 & -0.65 & 0.87\\
lat & 0.6 & -0.49 & -1.89 & 0.18 & 0.14 & 0.01 & 0 & 0.19 & 0.12 & 0 & -0.11 & 0.1\\
lon & 0.48 & 0.26 & -0.34 & 1.42 & 0.09 & 0 & -0.05 & 0.06 & 0.12 & 0 & 0 & 0.11\\
JJAtemp & 0.65 & 0.49 & 0 & 1.6 & 0.03 & 0 & 0 & 0 & 0.05 & 0 & 0 & 0\\
elev & 0.41 & 0.16 & -0.08 & 0.8 & 0.02 & 0 & 0 & 0 & 0.03 & 0 & 0 & 0\\
distSea & 0.23 & 0.02 & -0.29 & 0.43 & 0.03 & 0 & 0 & 0 & 0.06 & 0 & 0 & 0\\
MAP & 0.46 & 0.01 & -1.15 & 1.17 & 0.03 & 0 & 0 & 0 & 0.06 & 0 & 0 & 0\\
MSP & 0.9 & 0.96 & 0 & 2.05 & 0.03 & 0 & 0 & 0 & 0.05 & 0 & 0 & 0\\
wetDays & 0.3 & -0.01 & -0.63 & 0.57 & 0.04 & 0 & 0 & 0 & 0.07 & 0 & -0.01 & 0\\
JJAtemp.1 & 0.18 & 0.02 & -0.14 & 0.3 & 0.02 & 0 & 0 & 0 & 0.04 & 0 & 0 & 0\\
\hline
$\lambda$ & -- & 0.84 & 0.19 & 2.37 & -- & 7.08 & 4.01 & 11.74 & -- & 5.29 & 2.37 & 10.5\\
$\alpha$ & -- & 0.44 & 0.18 & 0.94 & -- & 4.17 & 2.37 & 7.02 & -- & 3.4 & 1.67 & 6.62\\
\hline\hline
\end{tabular}
\end{center}
\end{table}

The M-H proposal scheme outlined in Section~\ref{sec:modeling} and detailed in Appendix A was developed with the dual goal of eliminating the need for user-specified tuning parameters and the hope that by matching local curvature, acceptance probabilities would remain high.  Table~\ref{tab:acceptance} shows the acceptance probabilities for the MCMC chain run over the full data and indicates very high acceptance probabilities across the board.  The average acceptance probability for the random effects is well above .9 with the worst acceptance probability being a random effect for the shape parameter, at $.8$.  Likewise, the acceptance probabilities for the Gaussian process term $\lambda$ is above .8 for all three models.  This indicates that the MCMC proposal constructed in our implementation offers a useful solution.  The algorithm automatically tunes the proposals to the local curvature in the posterior distribution and is able to achieve high acceptance probabilities while doing so.  There are many improvements that could be made (which we discuss in Section~\ref{sec:conclude}) but the convergence shown in Figure~\ref{fig:convergence} and the acceptance probabilities below suggest our implementation is effective at approximating the posterior distribution.
\begin{table}
\caption{Acceptance probabilities for M-H steps in the MCMC run over the full dataset.  This shows the acceptance probability for the $\lambda$ term, as well as the worst, the average and best acceptance probabilities for the random effects for each of the three linear models in the BHM}\label{tab:acceptance}
\begin{center}
\begin{tabular}{l c c c c}
\hline\hline
Model & $\lambda$ & Worst $\tau$ & Mean $\tau$ & Best $\tau$\\
\hline
Location ($\mu$) & 0.84 & 0.83 & 0.96 & 1\\
Precision ($\kappa$) & 0.82 & 0.92 & 0.97 & 1\\
Shape ($\xi$) & 0.82 & 0.8 & 0.94 & 1\\
\hline\hline
\end{tabular}
\end{center}
\end{table}

\begin{figure}[htbp]
\begin{center}
\includegraphics[width = \linewidth]{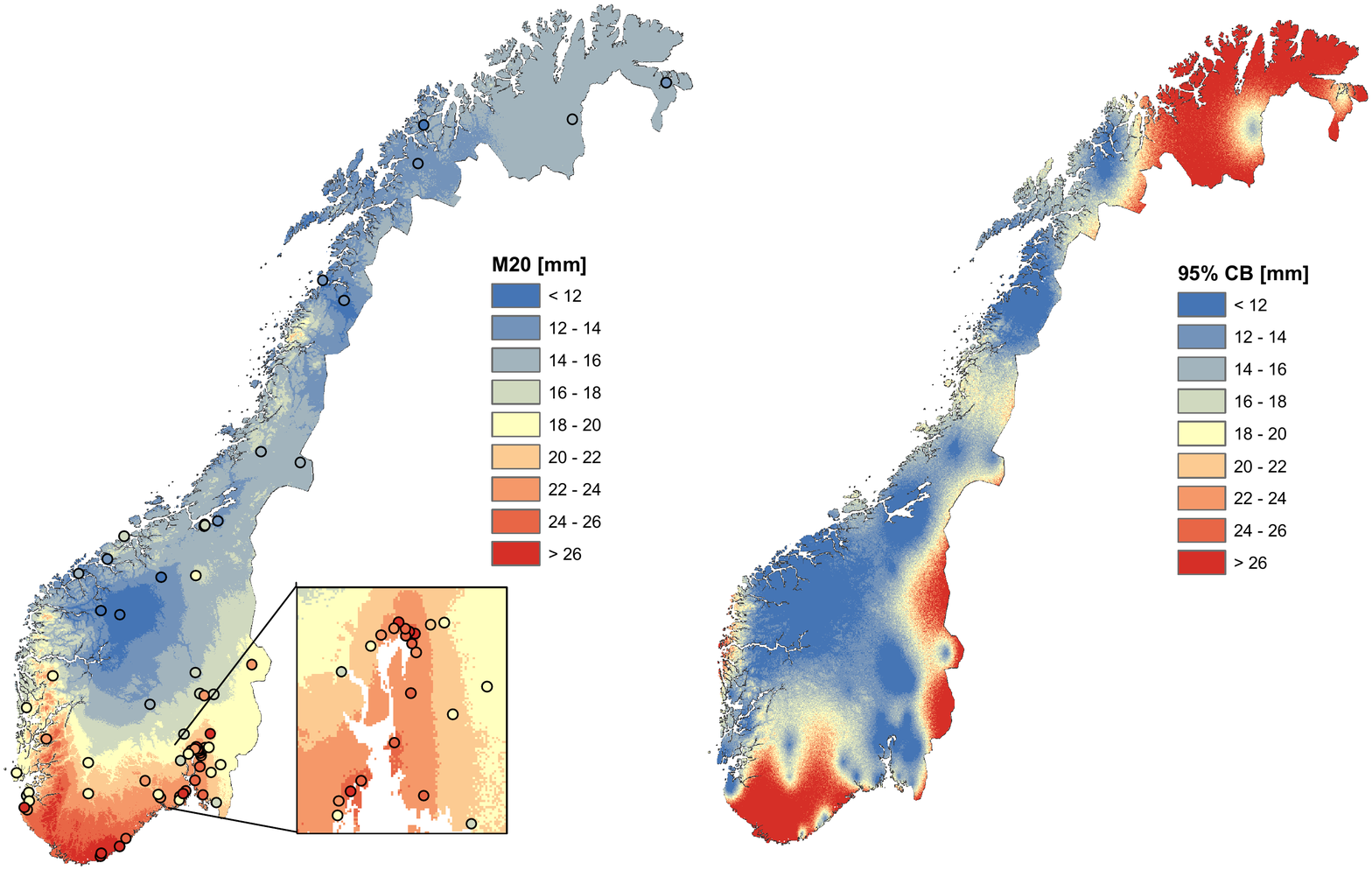} 
\caption[]{\label{fig:return_map}Left: Map of the modeled 20 year return level (M20) for hourly precipitation in Norway, estimated by the BMA approach. The dots refer to M20 estimated from a MLE fit to observations at the 69 locations. Right: The range of the 95\% confidence band for modeled M20.}
\includegraphics[width = 0.5\linewidth]{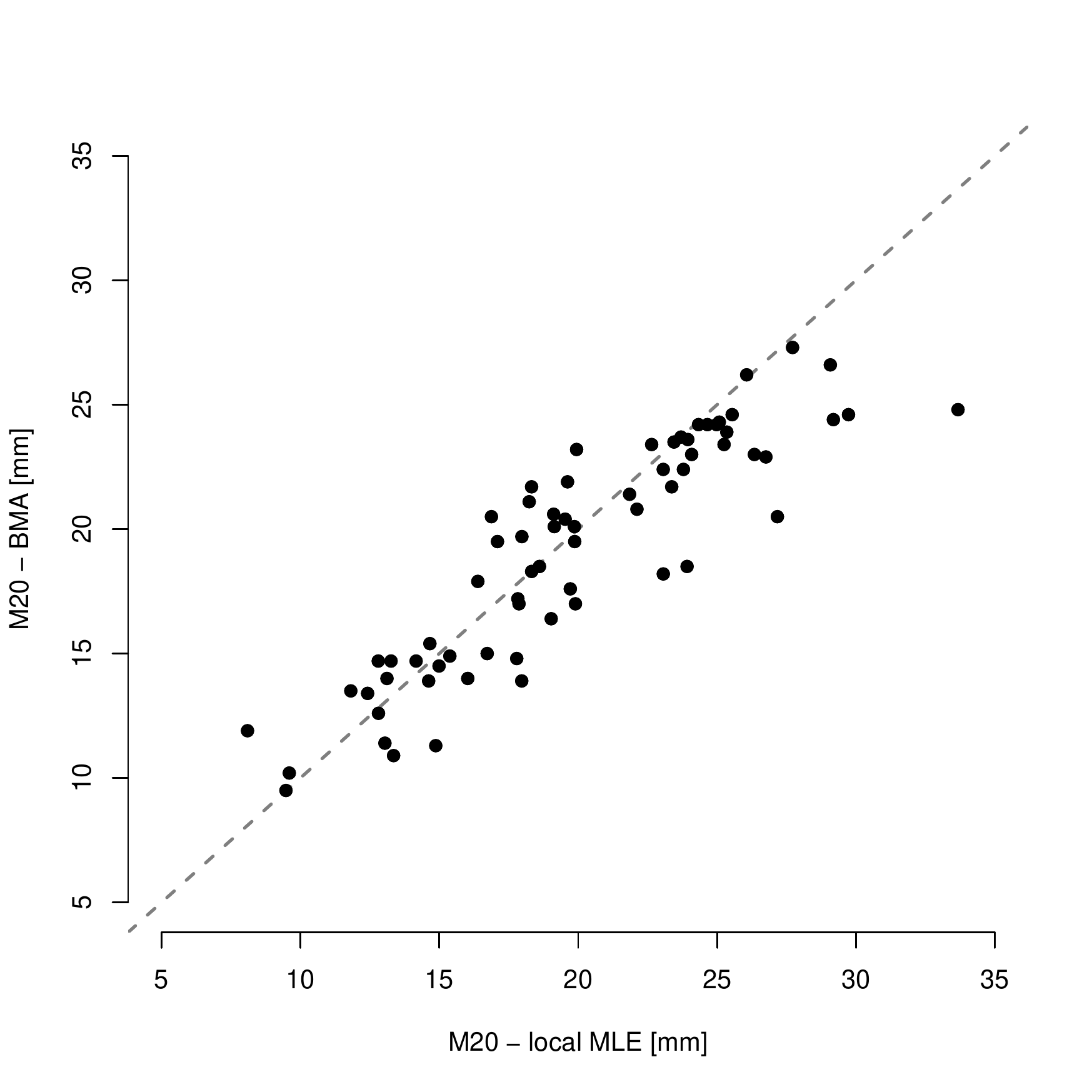} 
\caption[]{\label{fig:M20_scatter}Scatterplot at the 69 locations of modeled M20 for hourly precipitation estimated by the BMA approach versus a local MLE fit to observations.}
\end{center}
\end{figure}

The BMA run is finally used to construct return level maps over all of Norway, an example of which is shown in Figure~\ref{fig:return_map}. The map of estimated M20 reveals that BMA is able to reproduce reasonable values and a similar spatial pattern to what we expected. We have the largest values along the coast in the South, while the lowest values are seen in mountain regions and in the northern counties Nordland and Troms. \cite{MamenandIden2010} analyzed precipitation measurements of various durations in Norway and found that the largest return levels for hourly precipitation is seen in the southernmost coastal counties, including the Oslo-region. Relatively large values are also seen along the southwestern coast. We note that our model estimates somewhat lower values in the Oslo-region than in the southernmost regions. This is also reflected by a slight underestimation of the largest values in Figure~\ref{fig:M20_scatter}, where plot our BMA estimates against the local MLE's. We believe this is a feature of the relatively short observational series at many stations and the nature of intense showers. As the most extreme hourly events in this area are produced by small convective cells that hit locally, not all stations will experience them within their operative period, and the spread between single stations is larger. In constrast, in areas dominated by frontal and orographic precipitation the extreme values are more spatially consistent. 

The range of the confidence interval strongly depends on the number of stations nearby (cf. Fig.~\ref{fig:fig1}) and also on the magnitude of M20. In regions with very scarce station network (North-Norway and elevated areas), and where the terrain is more complex, the estimated return levels are subject to additional uncertainty both related to the gridding procedure in the covariates and the uncertain influence of these. We also recognize that there might be some correlation between the observational dataset and the covariates, as daily observations from the same locations go into the development of the gridded datasets. However, since we are using hourly observations these effects should be small.

\newpage

\begin{figure}[htbp]
\begin{center}
\subfigure[Empirical]{\includegraphics[width = .47\linewidth]{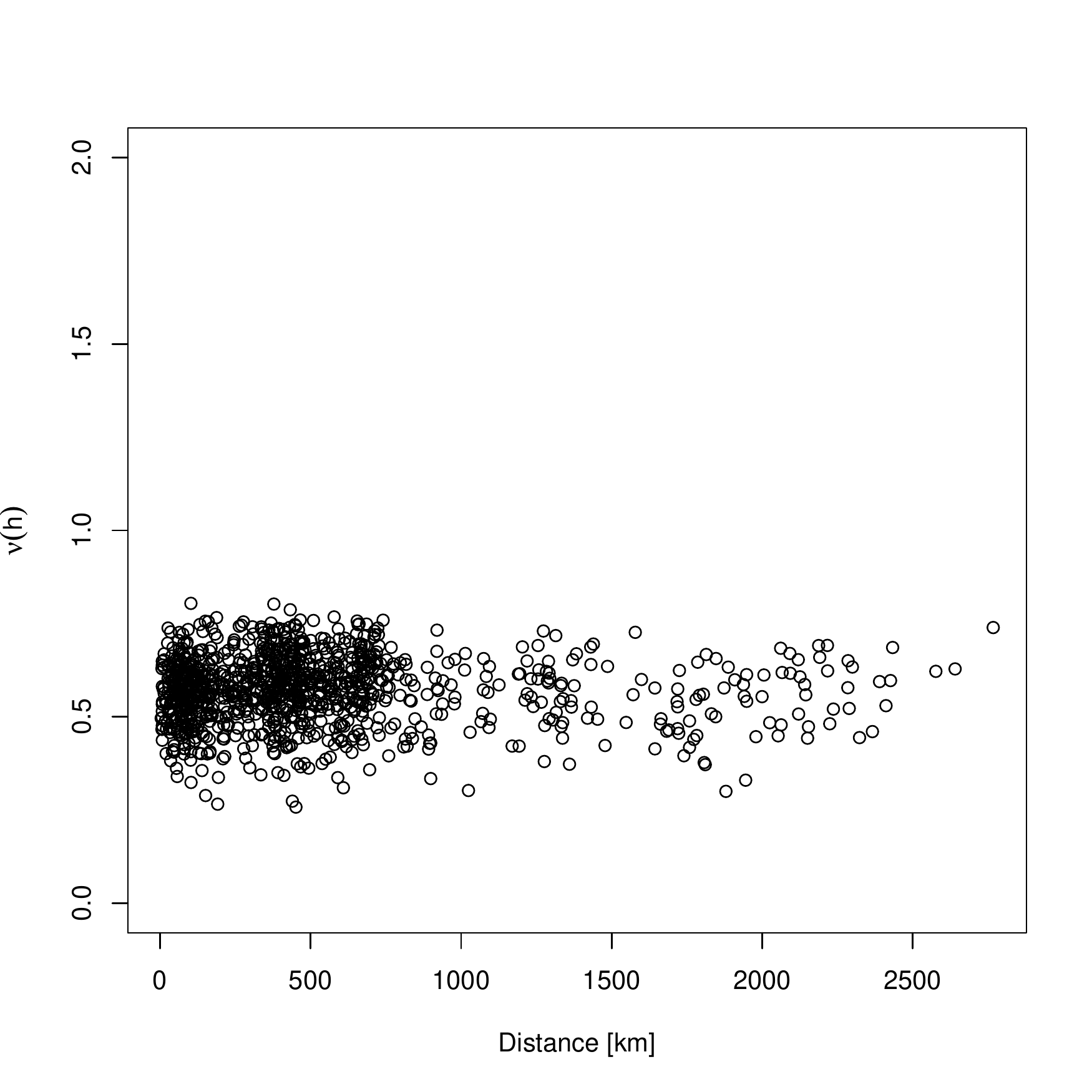}}
\subfigure[MLE]{\includegraphics[width = .47\linewidth]{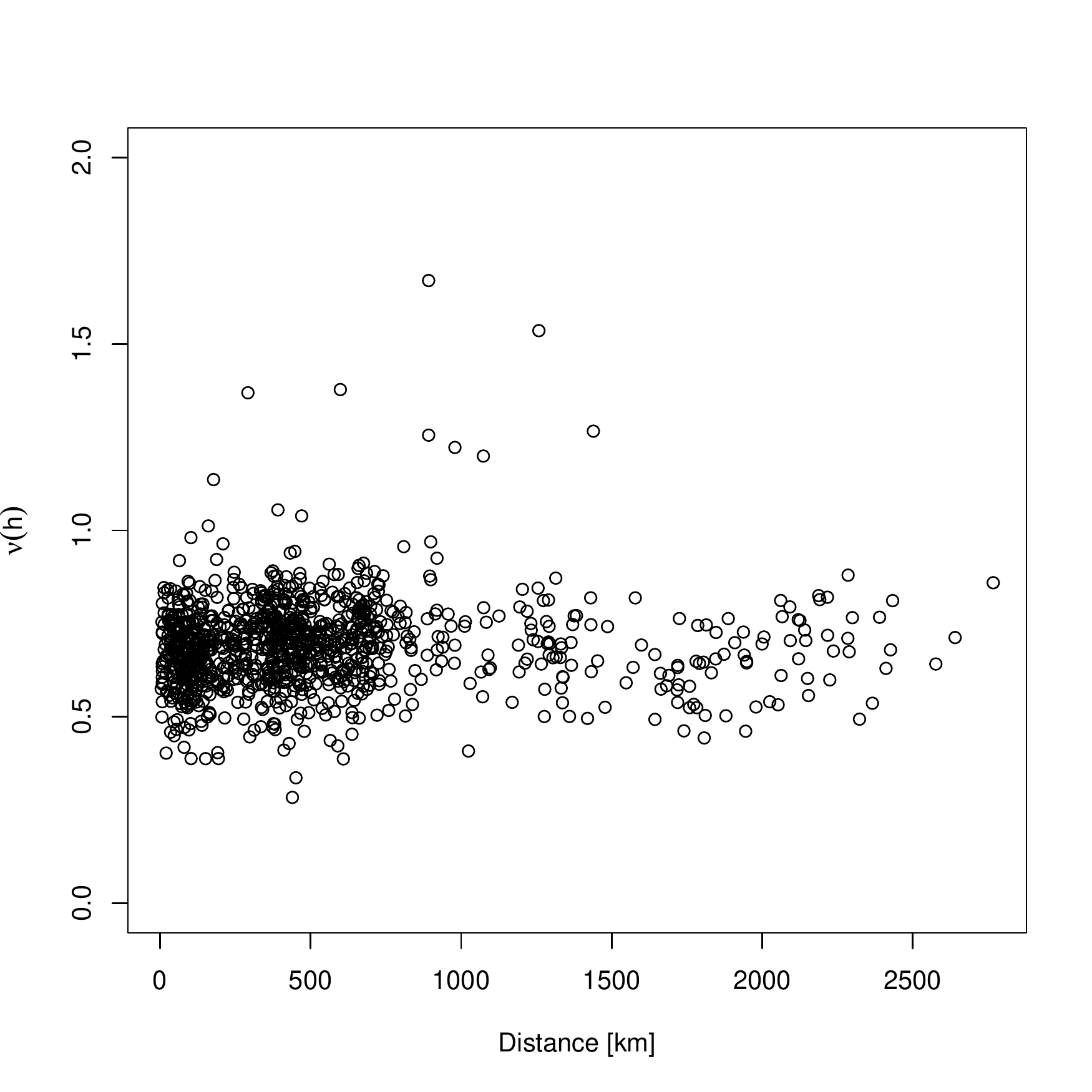}}
\caption{The madogram showing residual spatial dependence when using the empirical distribution for the marginal (left panel) and the MLE (right panel).}\label{fig:madogram}
\end{center}
\end{figure}

We finally conclude with a discussion of the conditional independence assumption.  The madogram \citep{cooley_et_2006} is a analogue to the variogram that assesses spatial dependence in extreme values.  Figure~\ref{fig:madogram} shows the madogram taken over our data where the marginal model used is either the empirical distribution (left panel) or the MLE (right panel).  As discussed in Section~\ref{sec:data}, individual sites exhibit missing information between years.  In order to obtain a sensible plot, pairs of observations are included only when they share 10 or more years of data.  In total, this still resulted in 1010 data points.

As we can see from Figure~\ref{fig:madogram} there appears to be little spatial dependence in observation pairs.  This appears to be due to the highly local behavior of extreme short-duration precipitation in Norway and furthermore suggests that the conditional independence assumption discussed in Section~\ref{sec:gp} appears reasonable for constructing marginal return levels estimates using the observation at hand.  Clearly, if observations were on a much finer scale, we would expect at that point to observe a higher degree of spatial residual dependence.

\section{Discussion}\label{sec:conclude}

We have developed a BHM for producing spatially continuous maps of return levels for hourly precipitation in Norway. The model spatially interpolates the GEV parameters estimated from observations via their relationship to geographical and meteorological variables on a fine grid. The inclusion of variable uncertainty was handled through BMA, in particular the use of conditional Bayes factors and M-H proposals were formed using Taylor-series expansions of posterior densities. This system was then shown to perform well at estimating return levels, both in terms of magnitude and spatial distribution, and represents an improvement on current methodology in Norway. As new and longer observational series become available, these can easily be incorporating, and the model can, with simple adjustments, be adapted to other durations and regions, given that a minimum amount of observations are available.   

Considerable work remains, both from the alghorithmic/methodological and application domains. While we have been happy with the present performance of the revised MCMC algorithm, block updating \citep{RueHeld2005} is a clear next step. Ultimately, incorporating concepts related to Riemannian manifold Hamiltonian sampling \citep{girolami_calderhead_2011}, should be entertained, especially as the observation network grows.  Furthermore, the current implementation is coded solely in {\tt R} and therefore exhibits a considerably slower run-time than other comparable methods that use lower-level programming languages.  A next step, once the algorithm is further developed methodologically, will be to rewrite the code base in e.g. {\tt C++}.

On a scientific level, it may be useful to consider segmenting Norway geographically to better address the various regimes present. In particular, a multiresolution approach to the Gaussian process could allow for the spatial over-dispersion to take on both global and local characteristics. As the quality of data improves, a peak-over-threshold approach, such as the GP model used in \cite{Cooleyetal2007}, could give more accurate estimation of local extremes. We would also like to test other covariates that might to a higher degree capture the spatial variability of the precision term $\kappa$ and shape $\xi$, such as one that more accurately separates areas dominated by convective and frontal precipitation and one that reflects orographic lifting. An obvious limitation in our model is the assumed stationarity in the covariates which leads to them competing over the degree of influence in different regions, thus an interesting next step would be to let the regression coefficients associated with the covariates vary in space. 

The latent variable approach applied here can reproduce the marginal behavior which is of main interest in infrastructure planning and support. However, it is important to note that the conditional independence assumption does not allow for estimation of single precipitation events since dependency between extremes at adjacent sites would not be modeled correctly. This means that while our model is able to capture climatological information at a given site, the total precipitation at this site at a specific time (in other words the weather) would most likely be under estimated.

\section{Acknowledgement}

The authors would like to thank the developer of the SpatialExtremes R-package, Mathieu Ribatet, for valuable answers to any question we had regarding the model, and Jostein Mamen at MET Norway for providing a ``clean'' observational dataset.
We also thank the Norwegian Water Resources and Energy Directorate (NVE), Norwegian railway authority (Jernbaneverket) and Norwegian public roads administration (Statens vegvesen) for providing financial support to Anita V. Dyrrdal's PhD project, and the NOTUR project for computational resources. The work of Alex Lenkoski and Thordis L. Thorarinsdottir is supported by Statistics for Innovation $(sfi)^2$, in Oslo.

\bibliography{anita}

\vspace{15mm}
\noindent*Available at http://met.no/Forskning/Publikasjoner/

\appendix

\section{MCMC updates of the random effects and the related hyperparameters}

Here, we discuss the MCMC updates of the Gaussian processes $\tau_s^\mu$, $\tau_s^\kappa$ and $\tau^\xi_s$ for each $s \in \mc{S}_o$ as well as the related hyperparameters $\alpha_\nu$ and $\lambda_\nu$ for $\nu \in \{ \mu, \kappa, \xi \}$. Most of these parameters require a Metropolis-Hastings update and the associated Hastings ratios \citep[e.g.][]{Hoff2009} can be calculated in a straight-forward manner.  That is, assume we want to update the parameter $\eta$ in our model, where $\eta$ is the current value. We then draw a new value $\eta'$ from a proposal distribution $pr(\eta' | \eta, \cdot)$ and accept the proposal with probability $\min \{ r, 1 \}$ where 
\bl\[
r = \frac{pr(\bs{y}| \eta', \cdot) pr(\eta' | \cdot) pr(\eta | \eta', \cdot)}{pr(\bs{y} | \eta, \cdot) pr (\eta| \cdot) pr(\eta' | \eta, \cdot)}.
\]\el
Here, $pr(\bs{y} | \eta, \cdot)$ denotes the likelihood of our full data set $\bs{y}$ which depends on $\eta$ and potentially other parameters which are kept fixed throughout, and $pr( \eta | \cdot)$ is the prior distribution of $\eta$ which similarly might depend on the other parts of the model.  Given the complexity of our model, it is vital to design efficient proposal distributions which return good proposals and are robust in that they do not require fine-tuning for each individual data set.  

\subsection{Random effects}

Under the Gaussian process model in \eqref{eq:GPmean} and \eqref{eq:GPcov}, the conditional distribution of $\tau_s$ (omitting the index $\nu$) conditional on the remaining values $\bs{\tau}_{\mc{S}_o \setminus s} = \{ \tau_{s'} \}_{s' \in \mc{S}_o \setminus s}$ is given by 
\bl\begin{equation}\label{eq:tauPrior}
\tau_s | \bs{\tau}_{\mc{S}_o \setminus s}, \alpha, \lambda \sim \mc{N} (\hat{\tau}_s, \varsigma_s),
\end{equation}\el
where 
\bl\begin{align*}
\hat{\tau}_s &= \mc{K}_{\alpha, \lambda}\big(s,\mc{S}_o\setminus s \big) \, \mc{K}^{-1}_{\alpha,\lambda}\big(\mc{S}_o\setminus s, \mc{S}_o\setminus s \big) \, \bs{\tau}_{\mc{S}_o\setminus s}\\
\varsigma_s &= \mc{K}_{\alpha, \lambda}\big(s,s\big) - \mc{K}_{\alpha, \lambda}\big(s,\mc{S}_o\setminus s \big) \, \mc{K}^{-1}_{\alpha,\lambda}\big(\mc{S}_o\setminus s, \mc{S}_o\setminus s\big) \, \mc{K}_{\alpha, \lambda}\big(\mc{S}_o\setminus s,s\big).
\end{align*}\el
We use this distribution as the prior distribution for $\tau_s$. 

For designing the proposal distribution, we employ the Gaussian approximation discussed, for instance, in Chapter 4.4 of \cite{RueHeld2005}.  Assume that the posterior distribution of the parameter $\tau'_s$ is written on the form
\bl\[
pr(\tau'_s| \cdot) \propto \exp \big( f(\tau'_s) \big),
\]\el
for some function $f$.  A quadratic Taylor expansion of the log-posterior $f(\tau'_s)$ around the value $\tau_s$ gives 
\bl\begin{align*}
f(\tau'_s) & \approx f(\tau_s) + f'(\tau_s) ( \tau'_s - \tau_s) + \frac{1}{2} f''(\tau_s) (\tau'_s - \tau_s)^2 \\
& = a + b \tau'_s - \frac{1}{2} c (\tau'_s)^2, 
\end{align*}\el
where $b = f'(\tau_s) - f''(\tau_s) \tau_s$ and $c = -f''(\tau_s)$.  The posterior distribution $pr(\tau'_s | \cdot)$ may now be approximated by 
\bl\[
\widetilde{pr}(\tau'_s | \cdot) \propto \exp \Big( -\frac{1}{2} c (\tau'_s)^2 + b \tau'_s \Big), 
\]\el 
the denisty of the Gaussian distribution $\mc{N}(b/c, c^{-1})$. We thus choose $\mc{N}(b/c, c^{-1})$ as our proposal distribution, where $\tau_s$ is the current state of the MCMC chain.  From \eqref{eq:tauPrior} it follows that 
\bl\begin{align*}
f'(\tau_s) &= \sum_{t = 1}^{T_s} \pderiv{}{\tau_s} \log pr(y_{ts}|\tau_s, \cdot) - \frac{\tau_s - \hat{\tau}_s}{\varsigma_s}\\
f''(\tau_s) &= \sum_{t = 1}^{T_s} \pderivsq{\tau_s} \log pr(y_{ts}|\tau_s, \cdot) - \frac{1}{\varsigma_s},
\end{align*}\el 
where $pr(y_{ts}|\tau_s, \cdot)$ is the GEV density in \eqref{eq:gev_density} and $T_s$ is the total number of observations available at location $s$.  

For the random effect in the location parameter $\mu$, we obtain 
\bl\begin{align*}
\pderiv{}{\tau_s^{\mu}} \log pr(y_{ts}|\tau_s^{\mu}, \cdot) & = (\xi_s + 1)\kappa_s h(y_{ts})^{-1} - \kappa_s h(y_{ts})^{-\xi^{-1}- 1} \\
\pderivsq{\tau_s^{\mu}} \log pr(y_{ts}|\tau_s^{\mu}, \cdot) &=\xi_s(\xi_s + 1)\kappa^2_sh(y_{ts})^{-2} - (\xi_s + 1)\kappa^2_sh(y_{ts})^{-\xi_s^{-1} - 2}.
\end{align*}\el
Let $\hat{\kappa}_s = \bs{x}_s^\top \bs{\theta}^\kappa$ denote the fixed effect in the inverse scale parameter at location $s$ and denote by $\epsilon_{ts} = y_{ts} - \mu_s$ the location residual at time $t$ and location $s$.  The derivatives with respect to the random effect in the inverse scale parameter $\kappa$ are then given by
\bl\begin{align*}
\pderiv{}{\tau_s^{\kappa}} \log pr(y_{ts}| \tau_s^\kappa, \cdot) &= \frac{1}{\hat{\kappa}_s + \tau_s^{\kappa}} - (\xi_s + 1)\epsilon_{ts}h(y_{ts})^{-1} + \epsilon_{ts}h(y_{ts})^{-\xi^{-1} - 1} \\
\pderivsq{\tau_s^{\kappa}} \log pr(y_{ts}|\tau_s^\kappa, \cdot) &=-\frac{1}{(\hat{\kappa}_s + \tau_s^{\sigma})^2} + (\xi_s + 1)\xi_s\epsilon_{ts}^2h(y_{ts})^{-2} - \epsilon_{ts}^2(\xi_s + 1)h(y_{ts})^{-\xi^{-1} - 2}
\end{align*}\el

The calculations for the shape parameter $\xi$ are somewhat more involved. Let $\hat{\xi}_s = \bs{x}_s^\top \bs{\theta}^\xi$ denote the fixed effect and set 
\bl\begin{align*}
f_1 &= \frac{\hat{\xi}_s + \tau_s^{\xi} + 1}{\hat{\xi}_s + \tau_s^{\xi}}\log h(y_{ts})\\
f_2 &= \exp\Big(-(\hat{\xi}_s + \tau_s^\xi)^{-1} \log h(y_{ts})\Big)
\end{align*}\el
We then obtain 
\bl\begin{align*}
\dot{f}_1 & = \pderiv{f_1}{\tau_s^{\xi}} = -\frac{\log h(y_{ts})}{(\hat{\xi}_s + \tau_s^{\xi})^2} + \frac{\hat{\xi} + \tau_s^{\xi} + 1}{\hat{\xi} + \tau_s^{\xi}}h(y_{ts})^{-1}\epsilon_{ts}\kappa_s\\
\dot{f}_2 & = \pderiv{}{\tau_s^\xi}f_2 = f_2  \left[\frac{\log h(y_{ts})}{(\hat{\xi}_s + \tau_s^\xi)^{2}} - \frac{h(y_{ts})^{-1}\kappa_s\epsilon_{ts}}{\hat{\xi}_s + \tau_s^{\xi}}\right],
\end{align*}\el
from which it follows that 
\bl\[
\pderiv{}{\tau_s^\xi} \log pr(y_{ts}| \tau_s^\xi, \cdot) = -\dot{f}_1 - \dot{f}_2.
\]\el
For the second derivative, similar calculations return
\bl\[
\pderivsq{\tau_s^\xi} \log pr(y_{ts}|\tau_s^\xi, \cdot) = \pderiv{}{\tau_s^{\xi}} \big(-\dot{f}_1 - \dot{f}_2\big) = g_1 - g_2 - g_3 + g_4,
\]\el
where
\bl\begin{align*}
g_1 &= -2(\hat{\xi}_s + \tau_s^{\xi})^{-3}\log h(y_{ts}) + (\hat{\xi}_s + \tau_s^\xi)^{-2}h(y_{ts})^{-1} \kappa_s\epsilon_{ts}\\
g_2 &= -\frac{h(y_{ts})^{-1}\epsilon_{ts}\kappa_s}{(\hat{\xi}_s + \tau_s^\xi)^2} - \frac{\hat{\xi}_s +\tau_s^\xi + 1}{\hat{\xi}_s + \tau_s^\xi}h^{-2}\epsilon_{ts}^2\kappa_s^2 \\
g_3&= \dot{f}_2\left[\frac{\log h(y_{ts})}{(\hat{\xi}_s + \tau_s^\xi)^{2}}\right] + f_2 \left[-2\frac{\log h(y_{ts})}{(\hat{\xi}_s + \tau_s^\xi)^{3}} + \frac{h(y_{ts})^{-1}\kappa_s\epsilon_{ts}}{(\hat{\xi}_s + \tau_s^\xi)^{2}}\right]\\
g_4 &= \dot{f}_2\left[\frac{h(y_{ts})^{-1}\kappa_s\epsilon_{ts}}{\hat{\xi}_s + \tau_s^{\xi}}\right] - f_2 \epsilon_{ts}\kappa_s\left[\frac{h(y_{ts})^{-1}}{(\hat{\xi}_s + \tau_s^{\xi})^{2}} + \frac{h(y_{ts})^{-2}\epsilon_{ts}\kappa_s}{\hat{\xi}_s + \tau_s^\xi}\right]. 
\end{align*}\el

\subsection{Hyperparameters}

Each Gaussian process prior has two hyperparameters $\alpha$ and $\lambda$ which determine the marginal variance and the range of the correlation in the random effects, respectively, see the model definition in \eqref{eq:GPcov}.  The updating steps for these parameters are the same for the three Gausssian processes, so we omit the index $\nu \in \{ \mu, \kappa, \xi\}$ in the following.  Let $\bs{E}(\lambda)$ be the $|\mc{S}_o| \times |\mc{S}_o|$ matrix where $[\bs{E}(\lambda)]_{ij} = \exp(-d_{ij}/\lambda)$ and thus $\mc{K}_{\alpha,\lambda}(\mc{S}_o, \mc{S}_o) = \alpha^{-1}\bs{E}(\lambda)$, and denote by $\bs{\tau} = \{ \tau_s\}_{s \in \mc{S}_o}$ the collection of $\tau_s$ at all locations $s$  in $\mc{S}_o$.  Assuming that the prior for $\alpha$ is of the form $\alpha \sim \Gamma(a_\alpha/2,b_\alpha/2)$, where the gamma distribution is parameterized in terms of shape and rate, simple calculations show that 
\bl\[
\alpha |\lambda, \bs{\tau} \sim \Gamma\left(\frac{|\mc{S}_0 + a_\alpha|}{2}, \frac{\bs{\tau}^\top E(\lambda)^{-1}\bs{\tau} + b_\alpha}{2}\right). 
\]\el
This parameter may therefore be sampled via a Gibbs step.

For the range parameter $\lambda$ we proceed in a similar manner as for the random effects above. However, the range parameter must fulfil $\lambda > 0$; our prior distribution is thus given by $\lambda \sim \Gamma (a_\lambda, b_\lambda)$ and we truncate the Gaussian proposal distribution at zero.  Let $\bs{D}$ be the $|\mc{S}_o| \times |\mc{S}_o|$ matrix such that $[\bs{D}]_{ij} = d_{ij}$.  We then have that
\bl $$
\log pr(\bs{\tau}|\alpha,\lambda, \bs{D}) \propto -\frac{\alpha}{2} \bs{\tau}^\top \bs{E}(\lambda)^{-1} \bs{\tau} + \frac{|\mc{S}_o|}{2} \log \alpha - \frac{1}{2} \log |\bs{E}(\lambda)|.
$$\el
To ease the notation, define
\bl\begin{align*}
\dot{\bs{E}}(\lambda) &= \frac{\partial}{\partial \lambda} \bs{E}(\lambda) = \frac{1}{\lambda^2}\bs{D} \circ \bs{E}(\lambda)\\
\ddot{\bs{E}}(\lambda)&= \pderiv{}{\lambda} \dot{\bs{E}}(\lambda) = -\frac{2}{\lambda^3}[\bs{D}\circ \bs{E}(\lambda)] + \frac{1}{\lambda^2}[\bs{D}\circ \dot{\bs{E}}(\lambda)],
\end{align*}\el
where $\circ$ denotes the Hadamard product.
Setting 
\bl\[
\bs{M}(\lambda) = \pderiv{}{\lambda} \bs{E}(\lambda)^{-1} = \bs{E}(\lambda)^{-1}[-\dot{\bs{E}}(\lambda)]\bs{E}(\lambda)^{-1},
\]\el
we have that
\bl\[
\pderiv{}{\lambda}\log pr(\bs{\tau}|\alpha, \lambda, \bs{D}) = - \frac{\alpha}{2} \bs{\tau}^\top \bs{M}(\lambda)\bs{\tau} -\frac{1}{2} tr\left\{\bs{E}^{-1}(\lambda)\dot{\bs{E}}(\lambda)\right\}. 
\]\el
Further calculations give
\bl\begin{align*}
\bs{N}(\lambda) & = \pderiv{}{\lambda}\bs{M}(\lambda) = \bs{M}(\lambda)[-\dot{\bs{E}}(\lambda)]\bs{E}(\lambda)^{-1} + \bs{E}(\lambda)^{-1}[-\ddot{\bs{E}}(\lambda)]\bs{E}(\lambda)^{-1}\\
& \qquad \qquad \qquad \, + \bs{E}(\lambda)^{-1}[-\dot{\bs{E}}(\lambda)]\bs{M}(\lambda) \\
tr\Big\{\bs{L}(\lambda)\Big\} & =  \pderiv{}{\lambda} tr\left\{ \bs{E}(\lambda)^{-1}\dot{\bs{E}}(\lambda)\right\} = tr\Big\{\bs{M}(\lambda)\dot{\bs{E}}(\lambda) + \bs{E}(\lambda)^{-1}\ddot{\bs{E}}(\lambda)\Big\}\\
\end{align*}\el
from which it follows that 
\bl\[
\pderivsq{\lambda} \log pr(\bs{\tau}|\alpha,\lambda, \bs{D}) = - \frac{\alpha}{2} \bs{\tau}'\bs{N}(\lambda)\bs{\tau} -\frac{1}{2}tr\{\bs{L}(\lambda)\} 
\]\el
These results, together with the derivatives of the $\Gamma (a_\lambda, b_\lambda)$ prior distribution then give 
\bl\begin{align*}
f'(\lambda) &= - \frac{\alpha}{2}\bs{\tau}'\bs{M}(\lambda)\bs{\tau} -\frac{1}{2}tr\Big\{\bs{E}(\lambda)^{-1}\dot{\bs{E}}(\lambda)\Big\}  - b_\lambda + (a_\lambda - 1)\lambda^{-1}\\
f''(\lambda) &=  - \frac{\alpha}{2} \bs{\tau}^\top\bs{N}(\lambda) \bs{\tau} -\frac{1}{2} tr\{\bs{L}(\lambda)\} - (a_\lambda - 1)\lambda^{-2}.
\end{align*}\el

\section{Prior Settings and Sensitivity}
We discuss the prior settings for our model and give some indication of the sensitivity of our results to these specifications.  For the parameters $\bs{\theta}^\kappa$ and $\bs{\theta}^{\xi}$ we chose standard normal priors $\mc{N}(\bs{0}, \mathbb{I}_p)$.  In point of fact, we saw almost no change in behavior at different settings of these priors.  The prior for $\bs{\theta}^{\mu}$, however, required a slight modification, where we set $\bs{\theta}_0 = (8,0,\dots,0)$.  We found that if the prior on the constant in this linear model was set to $0$, the lack of identification in the linear specification caused the random effects to occasionally become ``stuck'' at a mean value of $8$ while the constant term itself went to $0$. This is a clear issue with identification \citep[see][for a detailed study of these sorts of issues]{vandyk_meng_2001}.  Centering the prior for the constant term about $8$ is not only considerably more sensible in this application than $0$, it nullifies these identification issues.
Table~\ref{tab:gp_prior} shows the settings for the gamma distribution components for each of the Gaussian process parameters in each of the linear models.  These values were chosen based on ellicitation of the experience and inuition of the meterologists working on this project.
\begin{table}[htp]
\caption{Settings for the prior parameters of the Gaussian Process components used in our study}\label{tab:gp_prior}
\begin{center}
\begin{tabular}{l|cc|cc|cc|}
\hline\hline
Model&\multicolumn{2}{c|}{$\mu$}&\multicolumn{2}{c|}{$\kappa$}&\multicolumn{2}{c|}{$\xi$}\\
Parameter
& $\alpha$ & $\lambda$ & $\alpha$ & $\lambda$ & $\alpha$ & $\lambda$\\
\hline
a &  2 & 2 & 2 & 1.5 & 2 & 2\\
b & 6 & 2 & 2 & 1.5 & 1 & 1\\
\hline
\end{tabular}
\end{center}
\end{table}

After running the full model with these settings, tested the sensitivity to these settings.  We did this by running $24$ additional scenarios, were each hyper prior parameter was halved and doubled while holding all other parameters at the levels reported in Table~\ref{tab:gp_prior}.  Table~\ref{tab:gp_prior_median} shows the results of this study.  In each case, we report the median posterior value of the $\alpha$ and $\lambda$ variables for the model affected by a given alternative.  We note that techincally all models would affected by eacvh alternative.  In practice, spill-over effects to other models were minimal.

\begin{table}[htp]\caption{Posterior median of the $\alpha$ and $\lambda$ parameter for a given Gaussian process model when one of the associated prior parameters is altered by being}\label{tab:gp_prior_median}
\begin{center}
\begin{tabular}{l|cc|cc|cc|}
\hline\hline
Model&\multicolumn{2}{c|}{$\mu$}&\multicolumn{2}{c|}{$\kappa$}&\multicolumn{2}{c|}{$\xi$}\\
Scenario& $\alpha$ & $\lambda$ & $\alpha$ & $\lambda$ & $\alpha$ & $\lambda$\\
\hline
Base & 0.437 & 0.825 & 4.183 & 7.069 & 3.4 & 5.29\\
$a_\alpha$ Halved & 0.335 & 1.078 & 3.81 & 7.323 & 2.864 & 5.682\\
$a_\alpha$ Doubled & 0.614 & 0.606 & 4.691 & 6.755 & 4.122 & 4.912\\
$b_\alpha$ Halved & 0.579 & 0.732 & 7.094 & 5.723 & 4.141 & 4.827\\
$b_\alpha$ Doubled & 0.314 & 1.121 & 2.826 & 8.179 & 2.546 & 5.905\\
$a_\lambda$ Halved & 0.584 & 0.373 & 4.249 & 6.716 & 3.564 & 4.629\\
$a_\lambda$ Doubled & 0.349 & 1.515 & 4.044 & 7.813 & 3.106 & 6.733\\
$b_\lambda$ Halved & 0.319 & 1.716 & 3.498 & 11.542 & 2.876 & 8.538\\
$b_\lambda$ Doubled & 0.522 & 0.528 & 4.974 & 4.211 & 3.979 & 3.192\\
\hline\hline
\end{tabular}
\end{center}
\end{table}

\begin{figure}
\begin{center}
\subfigure[In Sample]{\includegraphics[width=.48\linewidth]{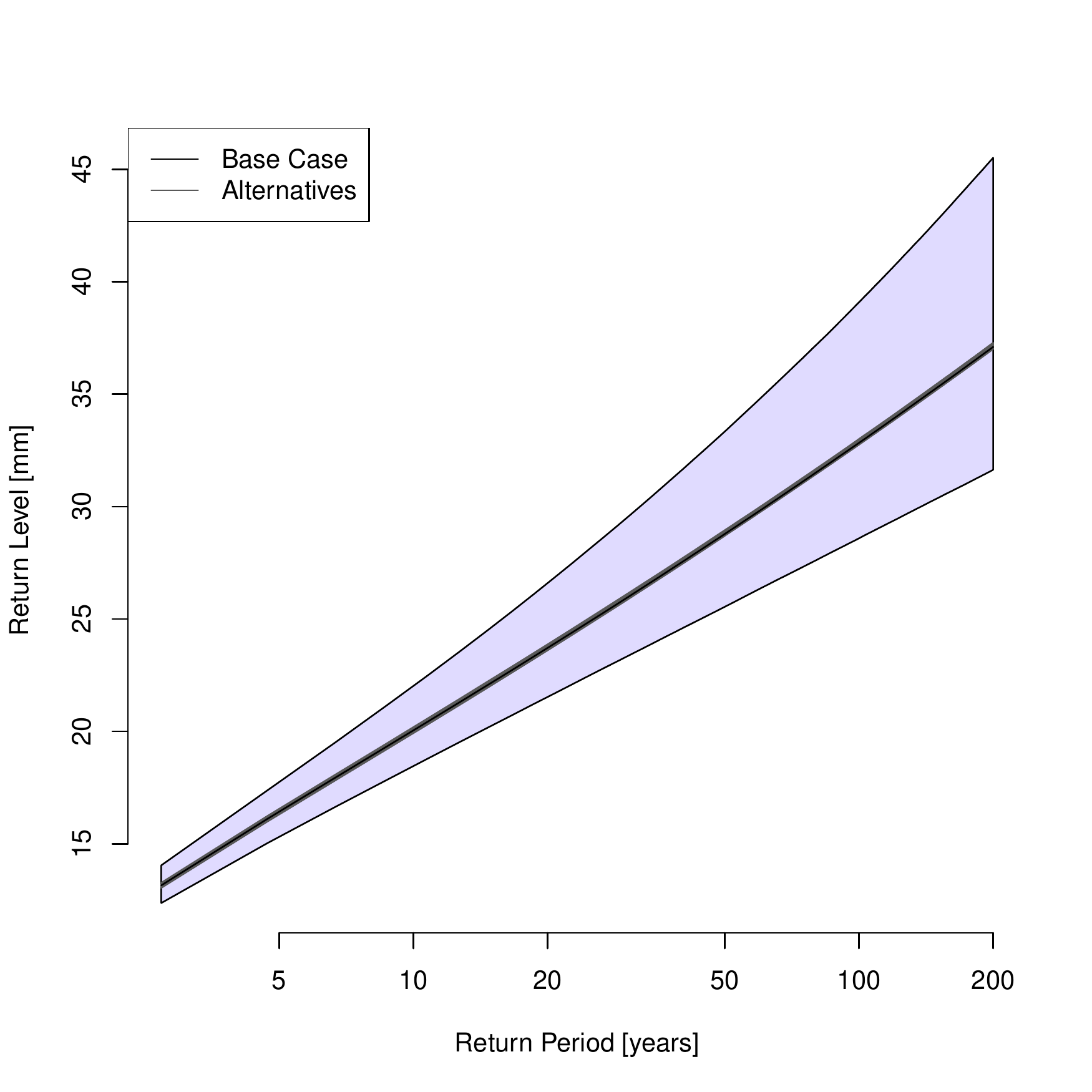}}
\subfigure[Out of Sample]{\includegraphics[width=.48\linewidth]{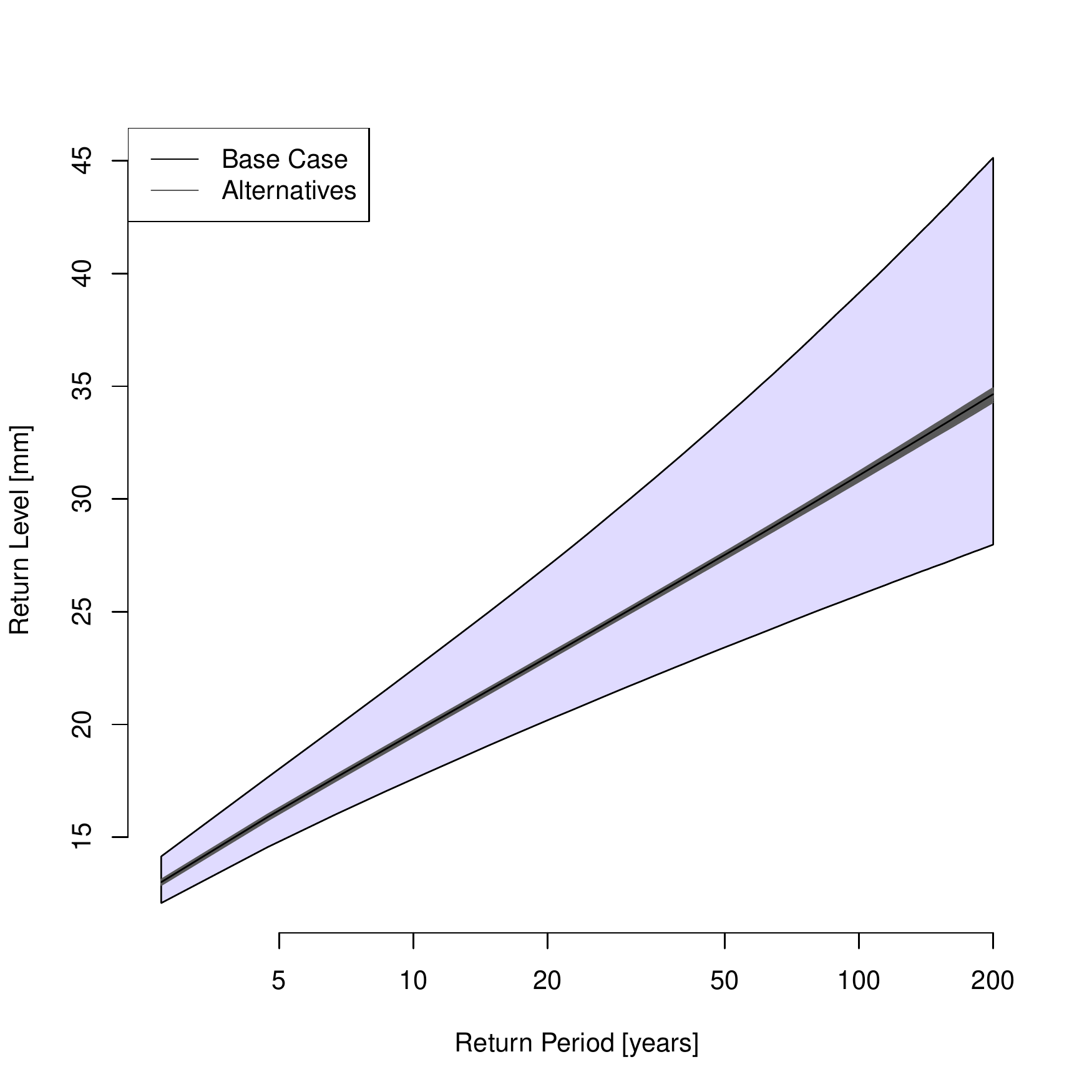}}
\caption{Posterior median return levels for Station 18701--both in and out of sample--under the base prior setting (black line) and alternative cases (grey lines) along with 90\% posterior interval under the base prior (shaded volume).}\label{fig:prior_sensitivity}
\end{center}
\end{figure}

Table~\ref{tab:gp_prior_median} shows that the posterior estimates of the hyperparameters are indeed affected by prior choices, in the directions that would be expected.  This is understandable, as hyperparameters are often sensitive to prior choice in hierarchical models.  However, Figure~\ref{fig:prior_sensitivity} shows that while the estimates of the hyperparameters are affected by prior choice, there is barely any concomitant effect on estimated return levels.  In Figure~\ref{fig:prior_sensitivity} we see the estimated return level for Station 18701. The black line shows the median estimate from the base prior choice, while the grey lines show medians from alternative choices.  For reference, the shaded volume shows the 90\% posterior interval for each return level under the base prior choice.  Finally, we considered both the case where Station 18701 is included during estimation, and when its return levels are estimated out of sample.

Figure~\ref{fig:prior_sensitivity} clearly shows that while the hyperparameter estimates are sensitive to prior settings, this has almost no subsequent effect on return level estimates.  The estimates for the out of sample study are naturally slightly more diffuse than for the in sample estimates, but these differences are minor in comparison to the overall statistical uncertainty in these values.  This indicates the prior values we have chosen are not having undue influence on our estimated return level maps.

\end{document}